\documentclass[preprintnumbers,showpacs,superscriptaddress,12pt,tightenlines,nofootinbib]{revtex4}

\usepackage{epsfig,latexsym,cancel,amssymb,amsmath}
\usepackage{mathtools}
\usepackage{graphicx}
\usepackage{feyn}
\usepackage{feynmp}
\usepackage[caption=false]{subfig}
\usepackage{epstopdf}
\usepackage{textcomp}
\usepackage{esint}
\usepackage[toc,page]{appendix}
\usepackage{dcolumn}
\usepackage{nohyperref}

\def\be{\begin{equation}}
\def\ee{\end{equation}}
\def\slash#1{\setbox0=\hbox{$#1$}#1\hskip-\wd0\dimen0=5pt\advance
       \dimen0 by-\ht0\advance\dimen0 by\dp0\lower0.5\dimen0\hbox
         to\wd0{\hss\sl/\/\hss}}

\newcommand{\noun}[1]{\textsc{#1}}
\newcommand{\eins}{\mbox{$1 \hspace{-1.0mm} {\bf l}$}}

\begin{document}

\preprint{CALT-TH-2017-005}
\preprint{ACFI-T17-21}

\title{Coherent $\mu-e$ Conversion at Next-to-Leading Order}

\author{Anthony Bartolotta$^{1}$}
\author{Michael J. Ramsey-Musolf$^{2,1}$}
\affiliation
{\hspace{5mm}
\\
$^{1}${\it Walter Burke Institute for Theoretical Physics, \\ California Institute of Technology\\
Pasadena, CA 91125 USA}
\\
$^{2}${\it Amherst Center for Fundamental Interactions\\
Department of Physics, University of Massachusetts Amherst\\
Amherst, MA 01003 USA}\\
}

\date{\today}

\begin{abstract}
We analyze next-to-leading order (NLO) corrections and uncertainties for coherent $\mu-e$ conversion . The analysis is general but numerical results focus on ${}^{27}\textrm{Al}$, which will be used in the \noun{Mu2E} experiment. We obtain a simple expression for the branching ratio in terms of Wilson coefficients associated with possible physics beyond the Standard Model and a set of model-independent parameters determined solely by Standard Model dynamics. For scalar-mediated conversion, we find that NLO two-nucleon contributions can significantly decrease the branching ratio, potentially reducing the rate by as much as 50\%. The pion-nucleon $\sigma$-term and quark masses give the dominant sources of parametric uncertainty in this case. For vector-mediated conversion, the impact of NLO contributions is considerably less severe, while the present theoretical uncertainties are comparable to parametric uncertainties. 
\end{abstract}

\maketitle

\section{Introduction}
\label{sec:Introduction}

Despite its many successes, the Standard Model (SM) has several phenomenological and theoretical shortcomings. Phenomenologically, the Standard Model provides no explanation for cosmic matter-antimatter asymmetry, the relic density of cold dark matter, or the dark energy associated with cosmic acceleration. The observation of neutrino oscillations requires extending the SM to account for non-zero neutrino masses. Theoretically, the SM suffers from a hierarchy problem, does not explain the quantization of electric charge, and simply parameterizes the vast range of elementary fermion masses and an associated mixing between flavor and mass eigenstates. 

The flavor problem remains, indeed, one of the most vexing. In the charged lepton sector, the presence of flavor mixing among the light neutrinos implies non-vanishing, though unobservably small, rates for flavor non-conserving processes, such as $\mu\to e\gamma$. Scenarios for physics beyond the Standard Model (BSM), however, can allow for significantly larger rates for such processes. The observation of charged lepton flavor violation (CLFV) may, thus, point to one or more of these proposals and shed new light on the flavor problem. This possibility motivates several current and future CLFV searches, such as the MEG experiment at the Paul Scherrer Institute (PSI) that has recently placed a limit of $< 4.2 \cdot 10^{-13}$ on the branching ratio for $\mu\to e\gamma$ \cite{MEG2016}; the upcoming \noun{Mu2e} and \noun{COMET} experiments at Fermilab and J-PARC, respectively, which will search for CLFV through the process of coherent $\mu-e$ conversion in the presence of a nucleus \cite{mu2eCDR, CometCDR}; and the possible search for $\mu\to 3 e$ at PSI. For recent experimental and theoretical reviews, see Refs.~\cite{Bernstein:2013hba,Lindner:2016bgg}

In this study, we focus on the process of coherent $\mu-e$ conversion. The quantity of interest is the branching ratio
\be
\textrm{BR}(\mu - e)=\frac{\mu^{-}+A(Z,N)\rightarrow e^{-}+A(Z,N)}{\mu^{-}+A(Z,N)\rightarrow\nu_{\mu}+A(Z-1,N)} \; ,
\ee
where the denominator is the rate for muon capture on a nucleus with $Z$ protons and $N$ neutrons with $A=Z+N$.
The standard model branching ratio for this process is predicted to be of the order $\textrm{BR}(\mu - e)\thickapprox10^{-54}\;$\cite{Czarnecki2011, Czarnecki2016}. At present, the best experimental bounds are from the \noun{SINDRUM II} collaboration which has constrained $\textrm{BR}(\mu - e) < 7\cdot10^{-13}\;$\cite{Wintz1998, mu2eCDR}. The next generation experiments, \noun{Mu2e} and \noun{comet}, are expected to improve these bounds by roughly four orders of magnitude, $\textrm{BR}(\mu - e)\lesssim 5 \cdot 10^{-17}\;$\cite{mu2eCDR,CometCDR}. 

Previous studies of coherent conversion have focused on leading order processes and their uncertainties \cite{Kitano2002, Kitano2002Erratum, Cirigliano2009, Crivellin2014-MU2E}. The primary goal of this work is to extend the analysis of coherent conversion to include next-to-leading order (NLO) corrections and their uncertainties. We focus primarily on phenomenological, dimension six effective semileptonic operators that may induce this CLFV conversion process. The framework of $SU(2)$ Chiral Perturbation Theory (ChPT) can then be used to relate operators in the phenomenological CLFV Lagrangian written in terms of quarks to the hadronic degrees of freedom relevant for nuclear physics dynamics. As the momentum transfer scale in coherent conversion is set by the muon mass and because the nucleons have no net strangeness, one might expect $SU(2)$ ChPT to be adequate for present purposes. However, CLFV operators involving strange quarks will still contribute to the conversion process. To assess the possible quantitative impact of these operators, we include their leading order contributions via $SU(2)$ flavor singlet terms in the chiral Lagrangian.
Doing so is preferable to the use of full $SU(3)$ ChPT as it allows for better control of both theoretical uncertainties and uncertainties introduced by the low energy constants of the chiral Lagrangian as shown in Refs. \cite{Crivellin2014-MU2E, Crivellin2014-WIMP}. We find that the strange quark contributions are generally small compared to other theoretical and parametric uncertainties, as seen in Table \ref{tab:BR-Scalar}. Thus, the use of $SU(2)$ ChPT in this context should be robust.

The primary results of this investigation are given in Eqs.~(\ref{eq:brmaster},\ref{eq:ConvAmpl-Scalar},\ref{eq:ConvAmpl-Vector}) and Tables \ref{tab:BR-Scalar} and \ref{tab:BR-Vector}. We summarize these results here for convenience. The branching ratio for coherent conversion can be written as a sum of four separate amplitudes, one for each spin configuration of the system,
\be
\label{eq:brmaster}
\textrm{BR}(\mu - e)_{\mathcal{A}} = \left( \frac{v}{\Lambda}\right)^4 \left[ \left|\tau_{\mathcal{A}}^{(1)} \right|^2 + \left|\tau_{\mathcal{A}}^{(2)} \right|^2 + \left|\tau_{\mathcal{A}}^{(3)} \right|^2 + \left|\tau_{\mathcal{A}}^{(4)} \right|^2\right] . 
\ee
Here, $\mathcal{A}=S (V)$ indicates a scalar (vector)-mediated conversion process; $v=246$ GeV, is the Higgs vacuum expectation value (VEV); $\Lambda$ is the mass scale associated with the BSM CLFV dynamics; and  the indices $w\in\left\{1,2,3,4\right\}$  denote each unique configuration as defined in Appendix F.

Within each conversion amplitude, it is possible to separate all model-independent parameters from the Wilson coefficients of the specific CLFV theory. Doing so for the case of scalar-mediated conversion yields
\be
\begin{aligned}
\left|\tau_{S}^{(w)}\right|^2 = & \left| \alpha_{S,ud}^{(w)} \left( \frac{C^{S,L}_{u} \pm C^{S,R}_{u}}{2} \right) + \alpha_{S,ud}^{(w)} \left( \frac{C^{S,L}_{d} \pm C^{S,R}_{d}}{2} \right) \right. \\
& \left. + \alpha_{S,s}^{(w)} \left( \frac{C^{S,L}_{s} \pm C^{S,R}_{s}}{2} \right) + \alpha_{S,\Theta}^{(w)} \left( \frac{C^{S,L}_{\Theta} \pm C^{S,R}_{\Theta}}{2} \right) \right|^{2}  ,
\label{eq:ConvAmpl-Scalar}
\end{aligned}
\ee
where the $C^{S,L}_q$ ($C^{S,R}_q$) denote the Wilson coefficients for a scalar interaction involving a left- (right-) handed muon interacting with a light quark of flavor $q=(u,d,s)$ as defined in Eq.~\eqref{eq:CLFV-Lagrangian-quarks};  where $C^{S,L}_\Theta$ ($C^{S,R}_\Theta$) give the corresponding heavy quark contributions entering via the energy-momentum tensor; and where positive (negative) signs are used for $w\in\left\{1,3\right\}$ ($w\in\left\{2,4\right\}$). All model-independent parameters have been absorbed in the definitions of the $\alpha$'s. These parameters are defined in Appendix \ref{app:BR-Formula} and their numerical values are given in Table \ref{tab:BR-Scalar}. 

Important for this work are the relative magnitudes of the LO, NLO one-loop, and NLO two-nucleon contributions for the scalar-mediated amplitudes. Each contribution contains a common factor of
\be
\sqrt{\frac{m_\mu}{\omega_\mathrm{capt}}}\, \left(\frac{m_\mu}{4\pi v}\right)^2 = 0.5563 \pm 0.0005 \; ,
\ee
where $\omega_\mathrm{capt}$ is the muon capture rate.
For $u$- and $d$-quarks, the LO contribution is obtained from the pion-nucleon $\sigma$-term
\be
\label{eq:alphaLO}
\alpha_{S,ud}^{(1)}(\mathrm{LO}) = \sqrt{\frac{m_\mu}{\omega_\mathrm{capt}}}\, \left(\frac{m_\mu}{4\pi v}\right)^2\ \frac{\sigma_{\pi N}}{{2\hat m}}\left(I_{S,p}^{(1)}+I_{S,n}^{(1)}\right) = 65 \pm 11 \; ,
\ee
where ${\hat m}$ is the average of $u$- and $d$-quark current masses, $\sigma_{\pi N}$ is the pion-nucleon $\sigma$-term, and the $I_{S,N}^{(1)}$ are integrals involving the overlap of incoming and outgoing lepton wave functions with the distributions of nucleons $N$. 

The NLO one-loop contribution is given by 
\be
\label{eq:alphaNLOloop}
-\alpha_{S,ud}^{(1)}(\mathrm{NLO\ loop}) = \sqrt{\frac{m_\mu}{\omega_\mathrm{capt}}}\, \left(\frac{m_\mu}{4\pi v}\right)^2\ \left( \frac{3 B_0m_\pi \mathring{g}_A^2}{64\pi \mathring{f}_\pi^2}\right) \ \Delta_S^{(1)} = 2.71\pm 0.30\; ,
\ee
where $B_0=2.75\pm 0.11$ GeV normalizes the scalar source in the chiral Lagrangian (See Section \ref{sec:Chiral-Power-Counting} below); $m_\pi$ and $\mathring{f}_\pi$ are the pion mass and LO pion decay constant; $\mathring{g}_A$ is the LO nucleon axial coupling; and 
\be
\Delta_S^{(1)}= \left({\tilde I}_{S,p}^{(1)}+{\tilde I}_{S,n}^{(1)}\right)-\left(I_{S,p}^{(1)}+I_{S,n}^{(1)}\right)= 3.96\pm 0.39\; ,
\ee
with the ${\tilde I}_{S,N}^{(1)}$ denoting additional overlap contributions associated with the one-loop amplitudes. The latter depend on the momentum transfer $|{\vec q}|$ to the outgoing electron. The appearance of the difference between the ${\tilde I}_{S,N}^{(1)}$ and ${I}_{S,N}^{(1)}$ reflects the vanishing of the one-loop amplitudes in the $|{\vec q}|\to 0$ limit. Note that for finite $|{\vec q}|$, $\alpha_{S,ud}^{(1)}(\mathrm{NLO\ loop})$ is finite in the $m_q\to 0$ limit; the explicit $m_\pi$ appearing in the prefactor of Eq.~\eqref{eq:alphaNLOloop} is compensated by a $1/m_\pi$ in $\Delta_S^{(1)}$.

The NLO two-nucleon contribution generates a significantly larger correction, given by 
\be
\label{eq:alphaNLO2N}
-\alpha_{S,ud}^{(1)}(\mathrm{NLO\ NN}) = \sqrt{\frac{m_\mu}{\omega_\mathrm{capt}}}\, \left(\frac{m_\mu}{4\pi v}\right)^2\ \left( \frac{3 B_0 K_F \mathring{g}_A^2}{64\pi \mathring{f}_\pi^2}\right)\ f^{SI}_\mathrm{eff}  \ \left(I_{S,p}^{(1)}+I_{S,n}^{(1)}\right)= 18.8^{+1.6}_{-9.5} \; ,
\ee
where $K_F$ is the nuclear Fermi momentum and $f^{SI}_\mathrm{eff} = 1.05^{+0.07}_{-0.53}$ is obtained by performing a one-body Fermi Gas average of the two-nucleon amplitude over a spin- and isospin-symmetric core. Note that both the NLO loop and NLO two-nucleon contributions enter with an opposite sign compared to the LO amplitude, thereby reducing the sensitivity to the $C^{S,L}_q$. The impact of the two-nucleon term may be particularly severe, with a reduction of up to $\sim 25\%$ ($50\%$) of the LO amplitude (rate), although the uncertainty in that estimate is also significant. A similar decomposition applies to the relative magnitudes of the $\alpha_{S,ud}^{(w)}$. We discuss the details leading to these results in the subsequent sections of the paper.

In the case of vector-mediated CLFV, the conversion amplitudes are given by, 
\be
\left|\tau_{V}^{(w)}\right|^2 = \left| \alpha_{V,u}^{(w)} \left( \frac{C^{V,L}_{u} \pm C^{V,R}_{u}}{2} \right) + \alpha_{V,d}^{(w)} \left( \frac{C^{V,L}_{d} \pm C^{V,R}_{d}}{2} \right) \right|^{2} .
\label{eq:ConvAmpl-Vector}
\ee
Once again, the positive signs are used for $w\in\left\{1,3\right\}$ while the negative signs are used for $w\in\left\{2,4\right\}$. The model-independent $\alpha$'s are defined in Appendix \ref{app:BR-Formula} and their numerical values are given in Table \ref{tab:BR-Vector}. The coherent vector amplitudes receive no NLO contributions via either loops or two-nucleon amplitudes. In the latter instance, the result is well-known from the analysis of meson-exchange contributions to the nuclear electromagnetic current. The leading non-trivial corrections to the charge operator appear at NNLO, where as the three-current receives NLO contributions. The latter, however, is not a coherent operator, so we do not consider the analogous current for the vector-mediated conversion process.

\begin{table}[h]
\centering
\begin{tabular}{l|D{,}{}{6,6}|D{,}{\pm}{6,3}|D{,}{\pm}{5,4}|D{,}{}{5,3}}
Parameter & \multicolumn{1}{l|}{Value} & \multicolumn{1}{l|}{LO Contribution} & \multicolumn{1}{l|}{NLO Loop} & \multicolumn{1}{l}{NLO Two-Nucleon} \\ \hline
$\alpha_{S,ud}^{(1)}$ & 43,{}^{+15}_{-12} & 65, 11 & -2.71,0.30 & -18.8,{}^{+9.5}_{-1.6}  \\
$\alpha_{S,s}^{(1)}$ & 3.71,\pm0.93 & 3.71, 0.93 & \multicolumn{1}{c|}{---} & \multicolumn{1}{c}{---} \\
$\alpha_{S,\Theta}^{(1)}$ & 8.43,\pm 0.13 & 8.43, 0.13 & \multicolumn{1}{c|}{---} & \multicolumn{1}{c}{---} \\ \hline
$\alpha_{S,ud}^{(2)}$ & 32,{}^{+11}_{-8} & 47.1, 8.3 & -1.96, 0.22 & -13.6,{}^{+6.9}_{-1.2} \\
$\alpha_{S,s}^{(2)}$ & 2.69,\pm 0.67 & 2.69, 0.67 & \multicolumn{1}{c|}{---} & \multicolumn{1}{c}{---} \\
$\alpha_{S,\Theta}^{(2)}$ & 6.11,\pm 0.10 & 6.11, 0.10 & \multicolumn{1}{c|}{---} & \multicolumn{1}{c}{---} \\ \hline
$\alpha_{S,ud}^{(3)}$ & -32,{}^{+8}_{-11} & -47.4, 8.3 & 1.96, 0.22 & 13.7, {}^{+1.2}_{-7.0} \\
$\alpha_{S,s}^{(3)}$ & -2.70,\pm 0.68 & -2.70, 0.68 & \multicolumn{1}{c|}{---} & \multicolumn{1}{c}{---} \\
$\alpha_{S,\Theta}^{(3)}$ & -6.15,\pm 0.10 & -6.15, 0.10 & \multicolumn{1}{c|}{---} & \multicolumn{1}{c}{---} \\ \hline
$\alpha_{S,ud}^{(4)}$ & -43,{}^{+12}_{-15} & -65, 11 & 2.68, 0.29 & 18.7,{}^{+1.6}_{-9.5} \\
$\alpha_{S,s}^{(4)}$ & -3.70,\pm 0.93 & -3.70, 0.93 & \multicolumn{1}{c|}{---} & \multicolumn{1}{c}{---} \\
$\alpha_{S,\Theta}^{(4)}$ & -8.41,\pm 0.13 & -8.41, 0.13 & \multicolumn{1}{c|}{---} & \multicolumn{1}{c}{---} \\ \hline
\end{tabular}
\caption{Table of branching ratio parameters for scalar-mediated conversion \label{tab:BR-Scalar}}
\end{table}

\begin{table}[h]
\centering
\begin{tabular}{l|D{,}{\pm}{6,4}}
Parameter & \multicolumn{1}{l}{Value} \\ \hline
$\alpha_{V,u}^{(1)}$ & 12.25, 0.13\\
$\alpha_{V,d}^{(1)}$ & 12.23, 0.27\\ \hline
$\alpha_{V,u}^{(2)}$ & -9.65, 0.11\\
$\alpha_{V,d}^{(2)}$ & -9.63, 0.21\\ \hline
$\alpha_{V,u}^{(3)}$ & -9.68, 0.11\\
$\alpha_{V,d}^{(3)}$ & -9.67, 0.21\\ \hline
$\alpha_{V,u}^{(4)}$ & 12.19, 0.13\\
$\alpha_{V,d}^{(4)}$ & 12.18, 0.27\\ \hline
\end{tabular}
\caption{Table of branching ratio parameters for vector-mediated conversion \label{tab:BR-Vector}}
\end{table}

Numerical results for the model-independent parameters $\alpha_{S,ud}^{(1)}$ {\em etc.} are given in Tables \ref{tab:BR-Scalar} and \ref{tab:BR-Vector}. As noted above, the NLO two-nucleon contributions may significantly degrade the sensitivity to the scalar-mediated interactions, whereas the vector-mediated sensitivities are unaffected to this order. We also note that the dominant sources of uncertainty in the scalar mediated branching ratio comes from the LO and NLO two-nucleon terms. The LO uncertainties are limited by the determination of the nucleon sigma-terms and quark masses. At NLO, the one-body Fermi Gas averaging of the two-nucleon term is the dominant source of uncertainty. This is again in contrast to the case of vector mediated conversion, for which the parametric and nuclear uncertainties are of the same order of magnitude as one expects for the NNLO contributions which are not explicitly computed in this work. 

This paper is organized as follows. In order to facilitate the reader's following the primary logic of our study, we relegate significant material to a number of Appendices that accompany the various sections. In Section \ref{sec:CLFV-Lagrangian}, we introduce the low-energy phenomenological effective CLFV Lagrangian  and discuss the corresponding Wilson coefficients. Section \ref{sec:Chiral-Power-Counting} and the accompanying Appendices \ref{app:ChiralLagrangian} and \ref{sec:LECs-and-LQCD} review the formalism of ChPT. We apply this framework to scalar-mediated CLFV in Section \ref{sec:Scalar-Mediated-Conversion}, deriving the LO and NLO matching of the  phenomenological CLFV operators onto the low-energy hadronic interactions at the one- and two-nucleon level. The one-body average of the two-nucleon interaction is discussed in Section \ref{sec:Effective-One-Nucleon} and Appendix \ref{app:OneBody}. In Section \ref{sec:Vector-Mediated-Conversion}, we consider the case of vector-mediated CLFV. The focus then turns to the sources of  theoretical hadronic uncertainties in Section \ref{sec:Hadronic} and Appendix \ref{app:Constants}. Section \ref{sec:Wavefunctions} discusses the calculation of the muon and electron wavefunctions, while Section \ref{sec:NuclearDensity} and Appendix \ref{app:Nuclear} examine uncertainties introduced by the nuclear density distributions. The branching ratio is calculated in Section \ref{sec:BranchingRatio} and the accompanying Appendices \ref{app:Overlap} and \ref{app:BR-Formula}, leading to our master formula in Eq.~\eqref{eq:brmaster}. The impact of the next-to-leading order corrections and uncertainties on the upcoming CLFV experiments is discussed in Section \ref{sec:Discussion}. We summarize our main results in Section \ref{sec:Conclusion} and provide Appendix \ref{app:BR-Formula} as a summary of how these results may be utilized.

%%%%%%%%%%%%%%%%%%%%%%%%%%%%%%%%%%%%%%%%%%%%%%%%%
\section{Quark-Level CLFV Lagrangian}
\label{sec:CLFV-Lagrangian}

There are a wide variety of extensions to the Standard Model that allow for CLFV. For an incomplete list of representative models, see, e.g. Refs. \cite{PatiSalam1974, Georgi1974, Farhi1981, Dugan1984, Randall1999, Branco2011, Bjorken1977, Buchmuller1987, McWilliams1980, Shanker1981, Dorsner2016}, and for more comprehensive surveys of the literature, see Refs. \cite{Kuno2001, Lindner:2016bgg}. Assuming that the process mediating CLFV occurs at a mass scale significantly greater than that of the momentum transfer involved in coherent $\mu-e$ conversion, $q_{T}^{2}\thickapprox m_{\mu}^{2}$, it suffices to concentrate on the low-energy effective Lagrangian which includes only SM fields as explicit degrees of freedom. 

In principle, one may start with an effective Lagrangian that respects the SU(3)$_C\times$SU(2)$_L\times$U(1)$_Y$ symmetry of the SM. Since our focus is on physics at the hadronic scale and below, we follow other authors \cite{Kitano2002, Kitano2002Erratum,Cirigliano2009} and work with an effective theory in which only the SU(3)$_C\times$U(1)$_\mathrm{EM}$ symmetry is manifest. The lowest dimension conversion operators of interest appear at mass dimension six:
\be
\begin{aligned}
\mathcal{L}_{\mathrm{CLFV}} = &
\displaystyle\sum_{f=\mathrm{u,d,s,c,b,t}} \frac{1}{\Lambda^{2}} \left[\lambda_{f}^{S,L}\bar{e}P_{L}\mu+\lambda_{f}^{S,R}\bar{e}P_{R}\mu + \mathrm{h.c.}\right]\bar{q}_{f}q_{f}\\
&+ \displaystyle\sum_{f=\mathrm{u,d,s,c,b,t}}\frac{1}{\Lambda^{2}} \left[\lambda_{f}^{V,L}\bar{e}\gamma^{\nu}P_{L}\mu + \lambda_{f}^{V,R}\bar{e}\gamma^{\nu}P_{R}\mu +\mathrm{h.c.}\right] \bar{q}_{f}\gamma_{\nu}q_{f}.
\end{aligned}
\label{eq:CLFV-Lagrangian-generic}
\ee
In principle, parity odd terms that couple to the pseudoscalar and axial-vector quark currents could be included, but this is not done as these contributions will be suppressed in coherent conversion. We also do not include the dipole operators relevant to $\mu\to e\gamma$ as their contributions to the coherent conversion process are typically suppressed relative to contributions from the scalar and vector interactions in Eq.~\eqref{eq:CLFV-Lagrangian-generic}.

In coherent conversion, the momentum transfer is roughly equal to the muon rest mass. As such, the dominant contributions from heavy quarks arise  through loop diagrams. Integrating out the heavy quarks results in an effective gluonic coupling that can be related to the stress energy tensor through the trace anomaly \cite{Shifman1978}. This procedure yields the Wilson coefficients
\begin{align}
C^{S,\mathcal{X}}_{f} &= \lambda^{S,\mathcal{X}}_{f} - \frac{2}{27} \sum_{h=\mathrm{c,b,t}} \frac{m_{f}}{m_{h}} \lambda^{S,\mathcal{X}}_{h}, \label{eq:eff-lep-coef-scalar}\\
C^{V,\mathcal{X}}_{f} &= \lambda^{V,\mathcal{X}}_{f}, \label{eq:eff-lep-coef-vector}\\
C^{\mathcal{X}}_{\Theta} &= \frac{2}{27} \sum_{h=\mathrm{c,b,t}} \frac{m_N}{m_{h}}\lambda^{S,\mathcal{X}}_{h} \label{eq:eff-lep-coef-mass},
\end{align}
where $m_N$ is the nucleon mass and $\mathcal{X}=L,R$ denotes the muon handedness. The resulting CLFV effective Lagrangian is
\be
\begin{aligned}
\mathcal{L}_{\mathrm{CLFV}} = &
\displaystyle\sum_{f=\mathrm{u,d,s}} \frac{1}{\Lambda^{2}} \left[ C^{S,L}_{f} \bar{e}P_{L}\mu + C^{S,R}_{f} \bar{e}P_{R}\mu + \mathrm{h.c.}\right] \bar{q}_{f}q_{f} \\
&+ \displaystyle\sum_{f=\mathrm{u,d,s}}\frac{1}{\Lambda^{2}} \left[ C^{V,L}_{f} \bar{e}\gamma^{\nu}P_{L}\mu + C^{V,R}_{f} \bar{e}\gamma^{\nu}P_{R}\mu + \mathrm{h.c.}\right] \bar{q}_{f}\gamma_{\nu}q_{f} \\
& + \frac{1}{M_N \Lambda^{2}} \left[ C^{L}_{\Theta} \bar{e}P_{L}\mu + C^{R}_{\Theta} \bar{e}P_{R}\mu + \mathrm{h.c.}\right] \Theta_{\mu}^{\mu} .
\end{aligned}
\label{eq:CLFV-Lagrangian-quarks}
\ee
For compactness of notation, we will define the effective CLFV currents
\begin{align}
J_{f} & = C^{S,L}_{f} \bar{e}P_{L}\mu + C^{S,R}_{f} \bar{e}P_{R}\mu + \mathrm{h.c.} , \label{eq:scalar-CLFV-current} \\
J_{f}^{\nu} &= C^{V,L}_{f} \bar{e}\gamma^{\nu}P_{L}\mu + C^{V,R}_{f} \bar{e}\gamma^{\nu}P_{R}\mu + \mathrm{h.c.} , \label{eq:vector-CLFV-current} \\
J_{\Theta} &= \frac{1}{M_N} \left[ C^{L}_{\Theta} \bar{e}P_{L}\mu + C^{R}_{\Theta} \bar{e}P_{R}\mu + \mathrm{h.c.}\right] , \label{eq:stress-energy-CLFV-current}
\end{align}
which couple to the quark scalar current, quark vector current, and trace of the stress energy tensor respectively. 

The Lagrangian in \eqref{eq:CLFV-Lagrangian-quarks} enables a model independent analysis of different theories with high-scale CLFV. However,  it will be used to describe CLFV processes involving light quarks  at the energy scales where QCD is non-perturbative and the relevant degrees of freedom are nucleons and mesons. The appropriate framework for doing this is ChPT.

%%%%%%%%%%%%%%%%%%%%%%%%%%%%%%%%%%%%%%%%%%%%%%%%%
\section{Chiral Power Counting and Chiral Lagrangians \label{sec:Chiral-Power-Counting} }

ChPT is the low-energy effective field theory of QCD \cite{Bernard1995}. At low energies QCD becomes confining which makes perturbative calculations with quarks and gluons intractable. Rather than using quarks and gluons as the fundamental degrees of freedom, ChPT replaces them with the bound states of mesons and baryons. Beyond these dynamical fields, ChPT can also include external source fields. These external sources will be used to incorporate the effective CLFV operators.

Starting from \eqref{eq:CLFV-Lagrangian-quarks}, one may use ChPT to relate the CLFV currents to an effective theory with multiple unknown LECs that must be matched onto experimental results. As is done in Appendix \ref{app:ChiralLagrangian}, it can be shown that these LECs are related to known nuclear matrix elements that appear in standard ChPT. The scalar and vector CLFV currents then appear in the chiral Lagrangian in an analogous manner to the quark mass and electromagnetic insertions respectively. However, as the CLFV currents do not scale with the quark mass they are assigned chiral order $\mathcal{O} \left( 1 \right)$. While $\mathcal{O} \left( 1 \right)$ in chiral power counting, the CLFV operators are still small in the sense that they correspond to high-scale physics and thus we may restrict our attention to terms with only a single CLFV insertion.

The inclusion of baryons in the chiral Lagrangian introduces additional complications in power counting beyond leading order. One well established method for dealing with these difficulties is Heavy Baryon Chiral Perturbation Theory (HBChPT) \cite{Jenkins1990}. This method requires a choice of reference velocity $V_{\mu}$ such that the decomposition of a nucleon's momentum, $P_{\mu}=m_{N}V_{\mu}+k_{\mu}$, yields a value of $k_{\mu}$ that is small compared to the chiral scale. For present purposes, the reference velocity is chosen to be $V_{\mu}=\left(1,0,0,0\right)$ in the rest-frame of the target nucleus. As a result, the magnitude of the residual three-momentum will be of the same order as the nuclear Fermi momentum, $|\vec{k}| \approx K_{F} \sim \mathcal{O}\left(q\right)$.

As noted in Section \ref{sec:CLFV-Lagrangian}, the momentum transfer scale for coherent conversion is set by the muon mass, $m_{\mu} \approx 106 \, \mathrm{MeV}$, which is comparable to the strange quark mass, $m_{s} \approx 92 \, \mathrm{MeV}$. Consequently, one should explicitly include the strange quark in the effective theory. On the other hand, the momentum transfer scale is not much greater than the strange quark mass and the nucleons have no net strangeness. Therefore, one might expect that the contributions of CLFV operators containing strange quark fields will be significantly smaller than the contributions of those coupling through the up and down quarks. If so, it may be advantageous to use $SU\left(2\right)$ ChPT with the leading order contributions of the strange quark operators treated as additional singlets under the flavor symmetry
rather than resorting to $SU\left(3\right)$ ChPT. As has been demonstrated previously \cite{Crivellin2014-MU2E, Crivellin2014-WIMP}, chiral $SU\left(2\right)$ allows for better control of both theoretical uncertainties and uncertainties introduced by the low-energy constants (LECs) of the chiral Lagrangian than is possible with chiral $SU\left(3\right)$. \textit{A priori}, the choice of chiral $SU \left( 2 \right)$ is not necessarily justified. However, the smallness of the strange quark contribution is borne out numerically in the results of Tables \ref{tab:BR-Scalar} and \ref{tab:BR-Vector}, justifying this approach.

Including the strange quark singlets in the chiral Lagrangian introduces an additional set of LECs that must be matched onto experimental results. The full set of relevant building blocks for the chiral Lagrangian and the complete chiral Lagrangian can be found in Appendix \ref{app:ChiralLagrangian}.

Lastly, chiral power counting for complete Feynman diagrams needs to be examined: in particular, how chiral power counting applies to multi-nucleon diagrams. One convenient power counting scheme only depends on the vertices and topological properties of the diagram \cite{Weinberg1991, Bedaque2002, Cirigliano2012}. An operator from the purely pionic sector $\mathcal{L}_{\pi\pi}^{(n)}$ is assigned the effective chiral power $\epsilon = n-2$ while operators from the pion-nucleon Lagrangian $\mathcal{L}_{\pi N}^{(n)}$ are given $\epsilon = n-1$. This effective chiral power is lower than the chiral order of the Lagrangian because the scaling of the propagators associated with a vertex must now be included with the vertex. This allows any diagram to be assigned an effective chiral order based on the following rule,
\be
\nu = 4 - A - 2C + 2L + \sum_{i}V_{i}\epsilon_{i} + \epsilon_{CLFV},
\label{eq:power-counting-scheme}
\ee
where $A$ is the number of external nucleons, $C$ is the number of connected parts of the diagram, $L$ is the number of loops, $V_{i}$ is the number of vertices with effective chiral power $\epsilon_{i}$, and $\epsilon_{CLFV}$ is the effective chiral power of the CLFV operator used.

%%%%%%%%%%%%%%%%%%%%%%%%%%%%%%%%%%%%%%%%%%%%%%%%%
\section{Scalar-Mediated Conversion}\label{sec:Scalar-Mediated-Conversion}

For the case of scalar-mediated conversion, the CLFV vector currents can be eliminated leaving the Lagrangian
\begin{gather}
\mathcal{L}_{\pi\pi}^{(0)} = \frac{\mathring{f}_{\pi}^{2}}{4} Tr\left[ \chi \left( U^{\dagger} + U \right) \right], \\
\mathcal{L}_{\pi N}^{(0)} =
\bar{N} \left[ \bar{c}_{5} \left( \chi \left( U + U^{\dagger} \right) - \frac{1}{2} Tr \left[ \chi \left( U + U^{\dagger} \right) \right] \right) + \bar{c}_{1} Tr \left[ \chi \left( U + U^{\dagger} \right) \right] + d_{1}^{S}\chi_{s} \right]N .
\end{gather}
The coefficients $\bar{c}_1$, $\bar{c}_5$, and $d_{1}^{S}$ are LECs that must be matched onto experimental data. The constant $\bar{c}_1$ is related to the nucleon mass in the isospin-symmetric limit, the constant $\bar{c}_5$ corresponds to the tree-level, isospin-breaking difference in the proton and neutron masses,  and $d_{1}^{S}$ is the strange quark contribution to the nucleon mass.

There are only two types of scalar insertion vertices that contribute at LO or NLO: insertion on a pion line from $\mathcal{L}_{\pi\pi}^{(0)}$ with effective chiral power $\epsilon_{CLFV} = -2$ and insertion on a nucleon line from $\mathcal{L}_{\pi N}^{(0)}$ with $\epsilon_{CLFV} = -1$. It should be stressed that these are the effective chiral powers used with the power counting scheme in \eqref{eq:power-counting-scheme} and do not correspond to how these terms in the Lagrangian scale with the power of $m_{\pi}$ or small momentum $q$. There are additional types of vertices at the same chiral order, but these will involve an even number of extra pions connected to the vertex; as such, these vertices can only contribute to diagrams at NNLO and beyond.

There are four possible diagrams that may contribute at LO and NLO. These can be divided into three categories: 

\begin{figure}[t]
	\hspace*{\fill}
	\subfloat[justification=justify][Leading order diagram consisting of a tree level insertion of a CLFV vertex. \label{fig:Feynman-Scalar-Tree}]{
	\begin{fmffile}{fgraphs1}
		\begin{fmfgraph*}(60,60)
		\fmfset{thick}{1.25}
		\fmfpen{thick}
		\fmfforce{(.05w,.2h)}{i1} 
		\fmfforce{(.95w,.2h)}{o1}
		\fmfforce{(.5w, .75h)}{t1}
		\fmf{fermion, label=\large$k_{1}$,label.dist=10,label.side=right}{i1,v1}
		\fmf{fermion, label=\large$k_{1}^{'}$,label.dist=7.5,label.side=right}{v1,o1}
		\fmf{photon, tension=0}{t1,v1}
		\fmfv{decoration.shape=circle, decoration.filled=shaded, decoration.size=10}{v1}
		\end{fmfgraph*}
	\end{fmffile}
	} \hfill
	\subfloat[Next-to-leading order diagram with a purely pionic loop and single nucleon. \label{fig:Feynman-Scalar-Loop1}]{
	\begin{fmffile}{fgraphs2} 
		\begin{fmfgraph*}(60,60)
		\fmfset{thick}{1.25}
		\fmfpen{thick}
		\fmfforce{(.05w,.2h)}{i1} 
		\fmfforce{(.95w,.2h)}{o1}
		\fmfforce{(.5w,.75h)}{t1}
		\fmf{fermion, tension=100, label=\large$k_{1}$,label.dist=10,label.side=right}{i1,v1}
		\fmf{fermion, tension=100, label=\large$k_{1}^{'}$,label.dist=7.5,label.side=right}{v1,o1}
		\fmf{scalar, left=1, tension=.5}{v1,v2,v1}
		\fmf{photon}{t1,v2}
		\fmfv{decoration.shape=circle, decoration.filled=shaded, decoration.size=10}{v2}
		\fmfdot{v1}
		\end{fmfgraph*}
	\end{fmffile}
	}
	\hspace*{\fill}
	\newline
	\hspace*{\fill}
	\subfloat[Next-to-leading order sunset diagram with an internal pion and single nucleon. \label{fig:Feynman-Scalar-Loop2}]{
	\begin{fmffile}{fgraphs3} 
		\begin{fmfgraph*}(60,60) 
		\fmfset{thick}{1.25}
		\fmfpen{thick}
		\fmfforce{(.1w,.2h)}{i1} 
		\fmfforce{(.9w,.2h)}{o1}
		\fmfforce{(.5w,.75h)}{t1}
		\fmf{fermion, tension=100, label=\large$k_{1}$,label.dist=10,label.side=right}{i1,v1}
		\fmf{fermion, tension=100}{v1,v2}
		\fmf{fermion, tension=100, label=\large$k_{1}^{'}$,label.dist=7.5,label.side=right}{v2,o1}
		\fmf{scalar, left=.5}{v1,v3,v2}
		\fmf{photon, tension=2}{t1,v3}
		\fmfv{decoration.shape=circle, decoration.filled=shaded, decoration.size=10}{v3}
		\fmfdot{v1,v2}
		\end{fmfgraph*}
	\end{fmffile}
	} \hfill
	\subfloat[Next-to-leading order diagram that involves the exchange of a pion between two nucleons. \label{fig:Feynman-Scalar-Two-Nucleon}]{
	\begin{fmffile}{fgraphs4}
		\begin{fmfgraph*}(60,60) 
		\fmfset{thick}{1.25}
		\fmfpen{thick}
		\fmfstraight
		\fmfforce{(.05w, .1h)}{i1}
		\fmfforce{(.05w, .5h)}{i2}
		\fmfforce{(.95w, .1h)}{o1}
		\fmfforce{(.95w, .5h)}{o2} 
		\fmfforce{(.5w, .8h)}{t1}
		\fmf{fermion, tension=100, label=\large$k_{2}$,label.dist=7.5, label.side=left}{i1,v1}
		\fmf{plain, tension=100}{v1,v4}
		\fmf{fermion, tension=100, label=\large$k_{2}^{'}$,label.dist=7.5, label.side=left}{v4,o1}
		\fmf{fermion, tension=100, label=\large$k_{1}$,label.dist=7.5, label.side=left}{i2,v2}
		\fmf{plain, tension=100}{v2,v5}
		\fmf{fermion, tension=100, label=\large$k_{1}^{'}$,label.dist=7.5, label.side=left}{v5,o2}
		\fmf{scalar}{v2,v3,v4}
		\fmf{photon, tension=0}{t1,v3}
		\fmfv{decoration.shape=circle, decoration.filled=shaded, decoration.size=10}{v3}
		\fmfdot{v2,v4}
		\end{fmfgraph*} 
	\end{fmffile}
	}
	\hspace*{\fill} \newline
	\caption{The set of Feynman diagrams that contribute to coherent $\mu-e$ conversion through NLO in a scalar-mediated model of CLFV. The fermionic and scalar lines correspond to nucleons and pions respectively. The shaded vertex represents an insertion of a CLFV operator. For diagrammatic simplicity, the leptonic line is not featured but would connect to the CLFV vertex.}
\end{figure}
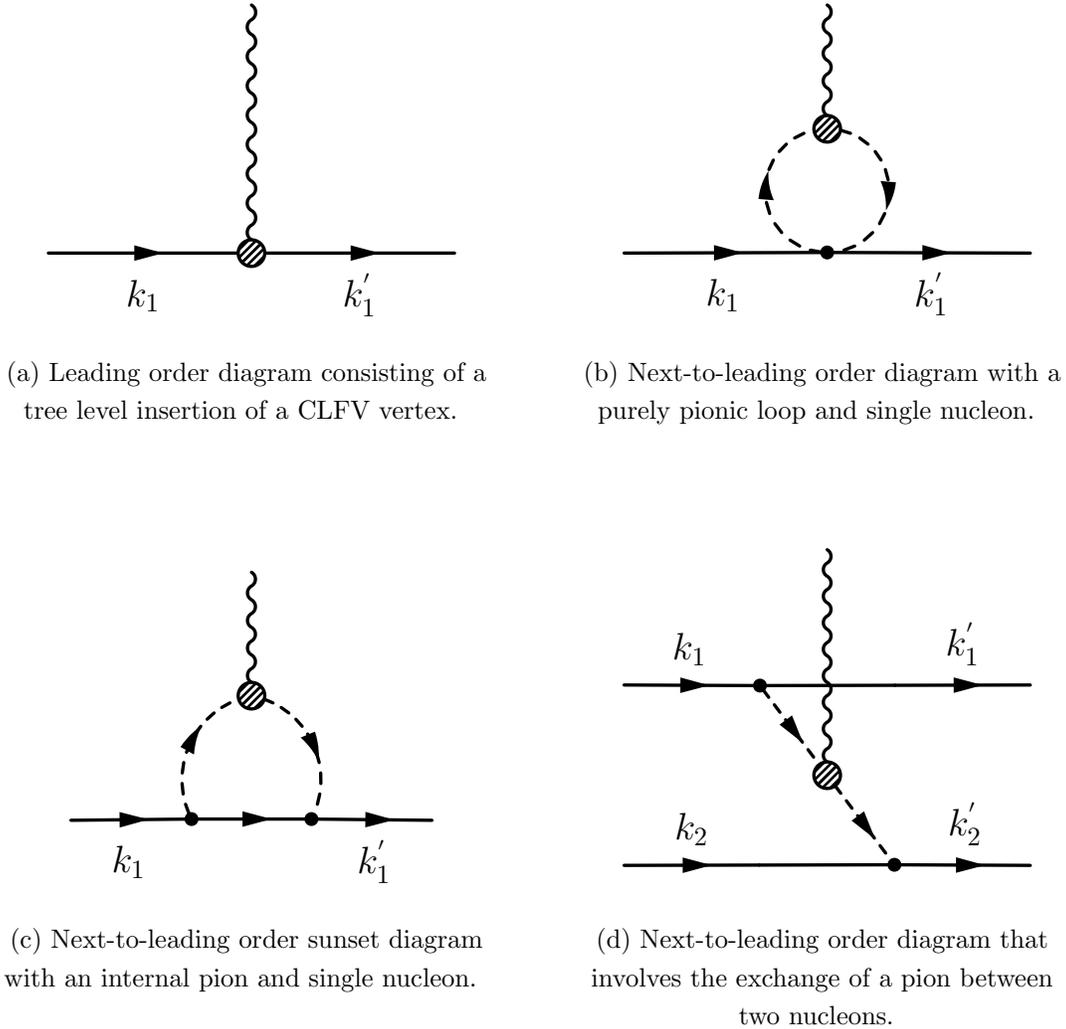

\begin{enumerate}
\item Single Nucleon, Tree-Level: The diagram of interest can be seen in Figure \ref{fig:Feynman-Scalar-Tree}. This consists of just the scalar insertion from $\mathcal{L}_{\pi N}^{(0)}$ on a single nucleon line that enters at effective chiral order $\nu =  3-3A$.

\item Single Nucleon, One-Loop: There are two possible diagrams that involve a pionic loop and a single nucleon. One diagram, Figure \ref{fig:Feynman-Scalar-Loop1}, consists of a single pion-nucleon vertex where the pion lines make a loop with the scalar insertion. The other, shown in Figure \ref{fig:Feynman-Scalar-Loop2}, is a sunset diagram with two pion-nucleon vertices where the scalar insertion happens on the internal pion line. Both of these diagrams involve the insertion of a CLFV operator from $\mathcal{L}_{\pi\pi}^{(0)}$ and enter at effective chiral order $\nu = 4-3A$. The diagram with a purely pionic loop, Figure \ref{fig:Feynman-Scalar-Loop1}, must vanish because the scalar insertion is symmetric in flavor indicies while the pion-nucleon vertex is anti-symmetric.

\item Two-Nucleon, Tree-Level: The diagram of interest can be seen in Figure \ref{fig:Feynman-Scalar-Two-Nucleon}. Two nucleons exchange a pion and the scalar insertion occurs on the internal pion line. The CLFV vertex is from $\mathcal{L}_{\pi\pi}^{(0)}$ and thus this diagram enters at effective chiral order $\nu = 4-3A$.
\end{enumerate}
It should be noted that these four diagrams have been analyzed previously in the context of dark matter direct-detection with an $SU(3)$ chiral Lagrangian \cite{Cirigliano2012}. The present formulation differs primarily in the use of an $SU(2)$ chiral Lagrangian to allow better control of uncertainties and a different treatment of the two-nucleon contribution. The present results were derived independently and agree with those of Ref. \cite{Cirigliano2012} in the limit of chiral $SU(2)$.

The diagrams involving only a single nucleon can be readily evaluated. Taken together, these three diagrams result in an effective nucleon-level CLFV Lagrangian,
\be
\label{eq:scalar-one-nucleon}
\begin{aligned}
\mathcal{L}_{1-N} =
\bar{N} &\left[ \left( 2 \bar{c}_{1} - \bar{c}_{5} \right) Tr\left[ \chi \right] + 2 \bar{c}_{5} \chi + d_{1}^{S} \chi_{S} + \frac{1}{\Lambda^{2}}J_{\theta} \right. \\
& \;\; \left. -\frac{3B_{0}m_{\pi}\mathring{g}_{A}^{2}}{64\pi \mathring{f}_{\pi}^{2} \Lambda^2} \left( J_{u} +J_{d} \right) \left( \frac{2+X_{\pi}}{\sqrt{X_{\pi}}} \textrm{arccot} \left( \frac{2}{\sqrt{X_{\pi}}} \right) - 1 \right) \right] N,
\end{aligned}
\ee
where the quantity $X_{\pi} = \left( \vec{q}_{T} \right)^{2} / m_{\pi}^{2}$ depends on the three momentum transferred to the nucleus, $\vec{q}_{T}$. The contribution from the stress-energy tensor has also been included in this effective Lagrangian..

The Lagrangian \eqref{eq:scalar-one-nucleon} can be further simplified by relating the LECs from $SU\left(2\right)$ ChPT to the contributions of the quark condensates to the proton and neutron mass. The difference between the proton and neutron mass is a NNLO effect that arises from isospin-symmetry breaking \cite{1993PhDT........44V, vanKolck1996}. Thus, we may take $\bar{c}_5 = 0$ as this is the LEC responsible for the mass splitting. The remaining LEC, $\bar{c}_1$, may be expressed at NLO accuracy in terms of $f_{u}^{N}$ ($f_{d}^{N}$), the fraction of the nucleon mass due to the $u$- ($d$-) quark condensate, as
\be
-4 B_0 \bar{c}_1 = \frac{m_N f_{u}^N}{m_{u}} = \frac{m_N f_{d}^N}{m_{d}} .
\label{eq:LEC-to-matrix-element}
\ee
Similarly, as shown in Appendix \ref{sec:LECs-and-LQCD}, the unknown LEC for the strange operator, $d_{1}^{S}$, can be matched onto the nucleon mass contribution from the strange quark condensate, $f_{s}^{N}$. Rewritting $\chi$ and $\chi_{S}$ in terms of the effective CLFV currents (see Appendix \ref{app:ChiralLagrangian}), one finds the effective Lagrangian
\be
\label{eq:scalar-one-nucleon-incomplete}
\begin{aligned}
\mathcal{L}_{1-N} = \frac{1}{\Lambda^{2}} \bar{N} &\left[ \frac{m_{N} f_{u}^{N}}{m_{u}} J_{u} + \frac{m_N f_{d}^{N}}{m_{d}}  J_{d} + \frac{m_N f_{s}^{N}}{m_{s}}  J_{s} + J_{\theta} \right. \\
& \;\; \left. - \frac{3B_{0}m_{\pi}\mathring{g}_{A}^{2}}{64\pi \mathring{f}_{\pi}^{2}}  \left( J_{u} +J_{d} \right) \left( \frac{2+X_{\pi}}{\sqrt{X_{\pi}}} \textrm{arccot} \left( \frac{2}{\sqrt{X_{\pi}}} \right) -1 \right)\right]N.
\end{aligned}
\ee

However, as we are working to NLO in $SU(2)$ ChPT, it is more appropriate to parameterize the effective Lagrangian in terms of the isospin-symmetry breaking parameter $\xi = \frac{m_d - m_u}{m_d + m_u}$. We will also introduce the isospin-symmetric quantities $\hat{m} = \frac{m_u + m_d}{2}$, the isospin averaged quark mass, and $\sigma_{\pi N}$, the pion-nucleon sigma-term. 

As has been shown previously in the literature \cite{Crivellin2014-MU2E, Crivellin2014-WIMP}, significant care must be taken to disentangle three flavor uncertainties when providing the chiral expansion for $f_u^N$ and $f_d^N$. These chiral expansions are known through NNLO \cite{Crivellin2014-WIMP}. As the present analysis of coherent $\mu-e$ conversion only extends to NLO, one finds
\be
\label{eq:quark-sigma}
m_N f_{q}^{N} = \frac{1}{2} \sigma_{\pi N} \left( 1\mp \xi \right).
\ee
In this expression, $q$ is a placeholder index for the $u$- ($d$-) quark condensate which is given by the negative (positive) sign. In terms of the isospin average quark mass, the $u$- ($d$-) quark mass is given by the negative (positive) sign in $m_q = \hat{m} \left( 1 \mp \xi \right)$. It is then straightforward to show using \eqref{eq:quark-sigma} that
\be
\label{eq:pion-nucleon-sigma}
\frac{m_N f_{q}^{N}}{m_q} = \frac{\sigma_{\pi N}}{2 \hat{m}}.
\ee
Making use of \eqref{eq:pion-nucleon-sigma}, one may rewrite \eqref{eq:scalar-one-nucleon-incomplete} to arrive at the final effective Lagrangian for the one-nucleon sector
\be
\label{eq:scalar-one-nucleon-LQCD}
\begin{aligned}
\mathcal{L}_{1-N} = \frac{1}{\Lambda^{2}} \bar{N} &\left[ \frac{\sigma_{\pi N}}{2 \hat{m}} \left( J_{u} + J_{d} \right) + \frac{\sigma_{s N}}{m_{s}}  J_{s} + J_{\theta} \right. \\
& \;\; \left. - \frac{3B_{0}m_{\pi}\mathring{g}_{A}^{2}}{64\pi \mathring{f}_{\pi}^{2}}  \left( J_{u} +J_{d} \right) \left( \frac{2+X_{\pi}}{\sqrt{X_{\pi}}} \textrm{arccot} \left( \frac{2}{\sqrt{X_{\pi}}} \right) -1 \right)\right]N.
\end{aligned}
\ee
In this expression, we have defined the strange-nucleon sigma-term $\sigma_{sN} = m_N f_{s}^N$.

The two-nucleon sector only includes a single tree-level diagram. This yields the effective two-nucleon Lagrangian
\be
\label{eq:two-nucleon-exact}
\mathcal{L}_{2-N} = -\frac{B_{0}\mathring{g}_{A}^{2}}{\mathring{f}_{\pi}^{2} \Lambda^{2}} \left( J_{u} +J_{d} \right) \frac{1}{\left( q_{1}^{2} - m_{\pi}^{2} \right) \left( q_{2}^{2} - m_{\pi}^{2} \right)} \sum_{a} \left( \bar{N}_{1}^{'}S\cdot q_{1}\tau_{a} N_{1} \right) \left( \bar{N}_{2}^{'}S\cdot q_{2}\tau_{a}N_{2} \right).
\ee
The quantities $q_{1}=k_{1}-k_{1}^{'}$ and $q_{2}=k_{2}-k_{2}^{'}$ are defined as the difference between the initial and final momenta of the two nucleons. The Lagrangians \eqref{eq:scalar-one-nucleon-LQCD} and \eqref{eq:two-nucleon-exact} closely mirror the results from the $SU\left(3\right)$ chiral Lagrangian \cite{Cirigliano2012}.

%%%%%%%%%%%%%%%%%%%%%%%%%%%%%%%%%%%%%%%%%%%%%%%%%
\section{Effective One Nucleon Operator}\label{sec:Effective-One-Nucleon}

The effective Lagrangian \eqref{eq:two-nucleon-exact} explicitly involves two external nucleons. Consequently, one requires the many-body wavefunctions for the initial and finial nuclei to calculate decay rates with this term. Carrying out such a complete, many-body computation goes beyond the scope of the present study. Nevertheless, in order to estimate the possible magnitude and relative sign of the two-nucleon contribution, we perform an average of the interaction over all core nucleons. In this approximation, it is assumed that every nucleon except for one valence nucleon is part of a spin-symmetric nuclear core. 
For the spatial wavefunction, the core nucleons can be approximated as being a degenerate Fermi gas. Such a distribution is fully characterized by its Fermi energy, $E_{F}$, or alternatively the Fermi momentum, $K_{F}$. For our purposes, it suffices to assume a common Fermi momentum for neutrons and protons. Isospin-breaking corrections should be of order $(N-Z)/A$. For earlier applications of this procedure to electroweak properties of nuclei, see, {\em e.g.}, Refs.~\cite{Haxton:1981sf,Anapole89,Musolf:1993mv}.

After making these approximations and summing over all contributions from the core nucleons, the spin-dependent and spin-independent parts of the resulting effective Lagrangian can be expressed in momentum space as
\begin{multline}
\label{eq:eff-one-nucleon-momentum}
\mathcal{L}_{\textrm{eff}} =  - \frac{3 B_{0}K_{F}\mathring{g}_{A}^{2}}{64 \pi (2\pi)^3 \mathring{f}_{\pi}^{2} \Lambda^{2}} \left( J_{u}+J_{d} \right) \\
\cdot \bar{N}(k_{f}) \left[ f^{SI}(\vec{q}_{T}, \vec{k}) \eins - f^{SD}(\vec{q}_{T}, \vec{k}) i \vec{\sigma} \cdot \left[\left(\frac{\vec{q}_T}{K_F}\right) \times \left(\frac{\vec{k}}{K_F}\right)\right]\right]  N(k_{i}),
\end{multline}
where the Pauli matrices are given by $\vec{\sigma}$, $\vec{k} = \frac{1}{2}\left(\vec{k}_i + \vec{k}_f\right)$ is the average of the initial and final nucleon three-momentum, and $\vec{q}_T = \vec{k}_f - \vec{k}_i$ is the three-momentum transferred to the nucleon. As this is an effective one-body operator, $\vec{q}_T$ is the same as the three-momentum transferred to the nucleus. The complicated dependence of the effective Lagrangian on $\vec{q}_T$ and $\vec{k}$ is encapsulated in the dimensionless functions $f^{SI}(\vec{q}_{T}, \vec{k})$ and $f^{SD}(\vec{q}_{T}, \vec{k})$. The full analytic forms of these functions are given in Appendix \ref{app:OneBody}.

For purposes of performing our numerical estimate, it is desirable to approximate these functions by constants. Doing so ensures that the effective Lagrangian \eqref{eq:eff-one-nucleon-momentum} remains local in position space, allowing seamless inclusion with \eqref{eq:scalar-one-nucleon-LQCD} as a single effective Lagrangian. As is demonstrated in Appendix \ref{app:OneBody}, the dimensionless functions $f^{SI}(\vec{q}_{T}, \vec{k})$ and $f^{SD}(\vec{q}_{T}, \vec{k})$ can be well approximated by the constants $f^{SI}_{\textrm{eff}}=1.05\pm0.07$ and $f^{SD}_{\textrm{eff}}=0.81\pm0.12$ respectively. The uncertainties in these constants include both the experimental uncertainties in the Fermi momentum of ${}^{27}\textrm{Al}$ and the anticipated errors induced by approximating the functions $f^{SI}(\vec{q}_{T}, \vec{k})$ and $f^{SD}(\vec{q}_{T}, \vec{k})$ by constants.

It is still necessary to include the errors induced by the core-averaging procedure itself. As a first pass, one may estimate these errors by examining previously studied cases in the literature where both core-averaged quantities and numerical many-body results were calculated. Analyzing previous results for the nuclear anapole moment \cite{Anapole89}, we infer that the core-averaging procedure may introduce an uncertainty of  $30\%$ to $50\%$ when the core is treated as a Fermi gas without short range correlations. It should also be noted that the core-averaged quantities generically over estimate the many-body contribution. Thus, we may conservatively take $f^{SI}_{\textrm{eff}}=1.05_{-0.53}^{+0.07}$ and $f^{SD}_{\textrm{eff}}=0.81_{-0.42}^{+0.15}$.

It is entirely possible, of course, that the results of a complete many-body computation would yield a result that falls outside of the aforementioned estimate. While the simplest single particle shell model description of $^{27}$Al is a $1d_{5/2}$ proton hole in $^{28}$Si, there is significant configuration mixing with two-particle excitations into the higher-lying $s_{1/2}$ and $d_{3/2}$ orbitals\footnote{We thank C. Johnson for a discussion of this point as well as for a numerical assessment using the Brown-Richter USDB interaction \cite{Brown:2006gx}.}. On the other hand, the results of elastic, magnetic electron scattering appear to agree well with the $1 d_{5/2}$ proton hole configuration description \cite{Donnelly:1984rg}\footnote{We thank T. W. Donnelly for alerting us to these results.}. Clearly, a detailed many-body computation using the two-body operator derived here will be needed for a definitive, quantitative assessment of the NLO two-body contribution.

As the conversion process is coherent, the spin-independent part of \eqref{eq:eff-one-nucleon-momentum} couples equally to all nucleons while the spin-dependent part is only relevant for unpaired nucleons. In the nuclear shell model, ${}^{27}$Al has only one unpaired proton. This results in a relative $1/A$ suppression of the spin-dependent term. This term may then be neglected as its contributions are comparable in magnitude to NNLO terms not considered in this analysis.

Returning to position space and combining this effective one-nucleon Lagrangian with the effective Lagrangian for the single nucleon sector yields the full effective Lagrangian for scalar-mediated conversion,
\begin{multline}
\label{eq:scalar-full-eff-Lagrangian}
\mathcal{L}_{\textrm{scalar}} = \frac{1}{\Lambda^{2}} \bar{N} \left[ \frac{\sigma_{\pi N}}{2 \hat{m}} \left( J_{u} + J_{d} \right) + \frac{\sigma_{s N}}{m_{s}}  J_{s} + J_{\theta} - \frac{3B_{0}K_{F}\mathring{g}_{A}^{2}}{64\pi \mathring{f}_{\pi}^{2}}  \left( J_{u} +J_{d} \right) f^{SI}_{\textrm{eff}} \right. \\
\left. - \frac{3B_{0}m_{\pi}\mathring{g}_{A}^{2}}{64\pi \mathring{f}_{\pi}^{2}}  \left( J_{u} +J_{d} \right) \left( \frac{2+X_{\pi}}{\sqrt{X_{\pi}}} \textrm{arccot} \left( \frac{2}{\sqrt{X_{\pi}}} \right) -1 \right) \right]N.
\end{multline}
We emphasize that the NLO loop and two-nucleon contributions enter with the opposite sign relative to the LO single nucleon terms, a feature reflected by the numerical results given in Table \ref{tab:BR-Scalar}.

%%%%%%%%%%%%%%%%%%%%%%%%%%%%%%%%%%%%%%%%%%%%%%%%%
\section{Vector-Mediated Conversion}\label{sec:Vector-Mediated-Conversion}

For the case of vector-mediated conversion, the scalar CLFV operators in the effective Lagrangian from Appendix \ref{app:ChiralLagrangian} can be removed. The vector CLFV operators enter the pion-nucleon Lagrangian at order $\mathcal{L}_{\pi N}^{(0)}$ but do not enter the purely pionic Lagrangian until $\mathcal{L}_{\pi\pi}^{(1)}$. This is because the vector CLFV current cannot couple to the scalar field except through a derivative. Thus pion loop and two-nucleon diagrams for vector-mediated CLFV will enter at NNLO instead of NLO as happened for scalar-mediated CLFV. Therefore, the only relevant diagrams will be tree-level insertions of the vector current. Replacing derivates with explicit factors of nucleon momentum, the CLFV Lagrangian may be rewritten as
\begin{multline}
\label{eq:vector-CLFV-momentum}
\mathcal{L}_{\textrm{vector}} = \bar{N}_{f} \left[  \left( V^{\mu} + \frac{\left( k_{f} + k_{i} \right)^{\mu}}{2M_{N}} - \frac{V\cdot \left( k_{f} + k_{i} \right)}{2M_{N}} V^{\mu} \right) \left( v + v^{(s)}\right)_{\mu} \right.\\
\left. - \frac{i}{M_{N}} \epsilon^{\mu \nu \rho \sigma} V_{\rho} S_{\sigma} \left( k_{f} - k_{i} \right) {}_{\mu} \left( \left( 1+\mathring{k}_{V} \right) v + \left( 1+\mathring{k_{s}} \right) v^{(s)} + \mu_{s} v_{s}^{(s)} \right)_{\nu} \right] N_{i},
\end{multline}
where we have used the relation $\left[ S^{\mu} , S^{\nu} \right] = i \epsilon^{\mu \nu \rho \sigma} V_{\rho} S_{\sigma}$, see Ref. \cite{Jenkins1990}, and identified the unknown LEC for the strange sector with the nucleon's strangeness magnetic moment, as demonstrated in Appendix \ref{sec:LECs-and-LQCD}.

The second set of terms that appear in the Lagrangian are spin-dependent while the first set are spin-independent. As discussed in Section \ref{sec:Effective-One-Nucleon}, the spin-dependent terms are suppressed by a factor of $1/A$ and it suffices to retain only the coherent, spin-independent terms. The Lagrangian also has terms of the form $V\cdot\left(\frac{k_{(i,f)}}{M_{N}}\right)$. As the external nucleons will be on shell, these terms are suppressed and actually enter at NNLO instead of NLO. Thus, these terms can be dropped from the effective nucleon CLFV Lagrangian leaving
\be
\mathcal{L_\textrm{vector}} =
\bar{N}_{f} \left[ \left( V + \frac{k_{f} + k_{i}}{2M_{N}} \right) \cdot \left( v + v^{(s)} \right) \right] N_{i}.
\ee
This Lagrangian depends not only on the magnitude of $\left( k_f + k_i \right)_\mu$ but also its direction. By parity-symmetry, the spatial components of $\left( k_f + k_i \right)_\mu$ must vanish but this still leaves the component $\left( k_f + k_i \right)_0$, the sum of nucleon kinetic energy. However, in the rest frame of the nucleus, this is equal to $V\cdot\left( k_f + k_i \right)$ which enters at NNLO as mentioned before. As a result, the final Lagrangian for vector-mediated conversion through NLO is just given by
\be
\label{eq:vector-full-eff-Lagrangian}
\mathcal{L_\textrm{vector}} = \bar{N} \left[ V\cdot \left( v + v^{(s)} \right) \right] N.
\ee

%%%%%%%%%%%%%%%%%%%%%%%%%%%%%%%%%%%%%%%%%%%%%%%%%
\section{Hadronic Uncertainties}
\label{sec:Hadronic}

The effective one-nucleon Lagrangians for scalar-mediated conversion, \eqref{eq:scalar-full-eff-Lagrangian}, and vector-mediated conversion, \eqref{eq:vector-full-eff-Lagrangian}, introduce a variety of physical parameters that must be matched onto experimental results. These include the light quark masses, pion decay constant, and nucleon axial-vector coupling, among others. 

The values of $\sigma_{\pi N}$, $\hat{m}$, and $\xi$ in addition to the other low energy parameters that appear in \eqref{eq:scalar-full-eff-Lagrangian} and \eqref{eq:vector-full-eff-Lagrangian} can be determined by making use of lattice QCD results. Modern $N_f = 2 + 1$ lattice QCD simulations provide realistic insight into several of these parameters with uncertainties that are smaller than their experimental counterparts. The low energy constants $\mathring{f}_\pi$ and $B_0$ along with the three light quark masses can be taken from the world average of lattice QCD results published by FLAG \cite{FLAG2016}. Similar world averages have been performed for both $\sigma_{s N}$ and $\sigma_{\pi N}$ \cite{Junnarkar2013,Alvarez-Ruso:2014}.

While lattice QCD simulations do provide better uncertainties for some quantities, others are best taken from experimental results. The pion and nucleon masses presented by the Particle Data Group are known to an exceptional degree of precision \cite{PDG2016}. Similarly, the nuclear axial-vector coupling, $g_A$, has been determined with high precision in ultra-cold neutron studies \cite{UCNA2010}. It should be stressed, however, that the experimentally observed values of the nucleon pole masses and nucleon axial-vector coupling are not quite the same as the objects that appear in the HBChPT Lagrangian. This is because the parameters in the HBChPT Lagrangian are the tree-level values taken in the chiral limit. Despite this difference, the experimental and chiral values only differ at NNLO and can thus be treated as equivalent for present purposes.

The full collection of low energy constants and their sources is summarized in Table \ref{tab:Constants} of Appendix \ref{app:Constants} along with the set of parameters that are derived from these constants.

%%%%%%%%%%%%%%%%%%%%%%%%%%%%%%%%%%%%%%%%%%%%%%%%%
\section{Wavefunctions of the Muon and Electron}
\label{sec:Wavefunctions}

Calculation of the coherent $\mu-e$ conversion rate requires knowledge of the wavefunctions for the bound muon and outgoing electron. Once captured by a nucleus, the muon relaxes to its ground state on a time scale much shorter than its mean lifetime. As such, one only needs to consider the captured muon in its ground state. The outgoing electron, however, is in a scattering state of fixed energy. These scattering states are highly relativistic as the electron receives nearly all of the decaying muon's energy, up to higher order corrections from nuclear recoil. To properly describe the wavefunction of the electron the Dirac equation must be used.

While the nucleus and electron or muon technically form a two-body system, reduced mass effects enter at NNLO and therefore the nucleus can be treated as a static source of a central potential. Following standard conventions \cite{Rose1961, Kitano2002, Kitano2002Erratum}, the time-independent Dirac equation in a spherically symmetric potential may be expressed as
\be
W\psi = \left[ -i\gamma_{5}\sigma_{r} \left( \partial_{r} + \frac{1}{r} - \frac{\beta}{r} K \right) + V \left( r \right) + m \beta \right] \psi,
\ee
where
\be
\begin{aligned}
\beta &= \left(
	\begin{array}{cc}
		\eins_{2} & 0\\
		0 & -\eins_{2}
	\end{array}
\right),
\gamma_5 = \left(
	\begin{array}{cc}
		0 & \eins_{2}\\
		\eins_{2} & 0
	\end{array}
\right),
\sigma_{r} = \left( 
	\begin{array}{cc}
		\hat{r}\cdot\vec{\sigma} & 0\\
		0 & \hat{r}\cdot\vec{\sigma}
	\end{array}
\right),\\
K &= \left( 
	\begin{array}{cc}
		\vec{\sigma}\cdot\vec{l}+\eins_2 & 0\\
		0 & -\left(\vec{\sigma}\cdot\vec{l}+\eins_2\right)
	\end{array}
\right).
\end{aligned}
\ee

In these expressions, the energy and mass of the particle are given by $W$ and $m$ respectively. The operator $K$ has been introduced for convenience as it commutes with the Hamiltonian while $\vec{\sigma}\cdot\vec{l}$ does not. This operator also has the useful property that $K^{2} = \hat{J}^{2} + \frac{1}{4}$. Letting $\kappa$ denote the eigenvalue of $K$ and $j\left(j+1\right)$ denote that of $\hat{J}^{2}$, it follows that $\kappa = \pm\left(j+\frac{1}{2}\right)$.

As the operators $J^{2}$, $J_{z}$, and $K$ commute with the Hamiltonian and each other, it is possible to work in a basis of states that have definite energy and eigenvalues for these operators. The two-component spinors in this basis will be denoted by $\chi_{\kappa}^{\mu} \left( \theta, \phi \right)$ where $\mu$ is the eigenvalue of $J_{z}$. This then allows the wavefunction to be decomposed as
\be
\psi = \left(\begin{array}{c}
g_{\kappa}\left(r\right)\chi_{\kappa}^{\mu} \left( \theta, \phi \right) \\
if_{\kappa}\left(r\right)\chi_{-\kappa}^{\mu} \left( \theta, \phi \right)
\end{array}\right),
\ee
where $g\left( r \right)$ and $f\left( r \right)$ are real valued functions. Expressed in terms of $g\left( r \right)$ and $f\left( r \right)$, the Dirac equation can be rewritten as the system of coupled differential equations
\be
\label{eq:Coupled-Dirac}
\frac{d}{dr} \left(
	\begin{array}{c}
		g\\
		f
	\end{array}
\right) = \left(
	\begin{array}{cc}
		-\frac{\kappa+1}{r} & W-V\left(r\right)+m\\
		-\left(W-V\left(r\right)-m\right) & \frac{\kappa-1}{r}
	\end{array}
\right) \left(
	\begin{array}{c}
		g\\
		f
	\end{array}
\right).
\ee
These coupled equations can then be solved numerically using the shoot-and-match procedure \cite{Silbar2011}.

As the muon is in its ground state, its wavefunction is normalized using the usual scheme
\be
\int d^{3}x\psi_{\kappa',\mu'}^{\left(\mu\right) \dagger}\left(x\right) \psi^{\left(\mu\right)}_{\kappa,\mu} \left(x\right) = \delta_{\mu',\mu} \, \delta_{\kappa',\kappa}.
\ee
The electron, however, is described by a scattering state which require a different normalization scheme. Because the wavefunction takes continuous energy eigenvalues these states are normalized as
\be
\int d^{3}x\psi_{\kappa',\mu',E'}^{\left(e\right) \dagger}\left(x\right) \psi^{\left(e\right)}_{\kappa,\mu,E} \left(x\right) = 2\pi\delta\left(E'-E\right)\delta_{\mu',\mu} \, \delta_{\kappa',\kappa}.
\ee

%%%%%%%%%%%%%%%%%%%%%%%%%%%%%%%%%%%%%%%%%%%%%%%%%
\section{Nuclear Density Distributions}
\label{sec:NuclearDensity}

Beyond the wavefunctions of the muon and electron, it is also necessary to determine the distribution of protons and neutrons in the nucleus of ${}^{27}\textrm{Al}$. These distributions directly enter the calculation of the decay rate and the proton density distribution indirectly impacts the muon and electron wavefunctions by virtue of determining the electric potential in the vicinity of the nucleus.

As the proton is electrically charged, its nuclear density distributions have been thoroughly explored through electron scattering experiments \cite{Negele1989}. These experiments have determined the nuclear charge density distribution of many nuclei to high precision in a model-independent manner \cite{deVries1987}. One such model-independent decomposition of the nuclear charge density distribution is the Fourier-Bessel expansion. Using this expansion, the distribution is given by the piecewise function 
\be
\label{eq:FourierBessel}
\rho_{p}\left(r\right) = 
\begin{cases}
\sum_{n} a_{n} j_{0}\left(\frac{n \pi r}{R}\right) & r \leq R\\
0 & r > R
\end{cases}.
\ee
There are a variety of ways to normalize this distribution, though the scheme $\int 4\pi r^{2} \rho\left(r\right)dr = Z$ will be used here. In \eqref{eq:FourierBessel}, the parameter $R$ acts as a cutoff radius for the charge and the set of parameters $a_n$ correspond to independent components of the charge density distribution. While the distribution is cut off at $r=R$, the distribution is defined such that it goes to zero in a continuous manner. The experimentally determined values of these parameters for ${}^{27}\textrm{Al}$ are given in Table \ref{tab:ProtonConst} of Appendix \ref{app:Nuclear}.

While the Fourier-Bessel parameters of Table \ref{tab:ProtonConst} are given without individual uncertainties, the uncertainty in the root-mean-square charge radius is known. Experimentally, $\left\langle r^{2} \right\rangle^{1/2}_{p} = 3.035\pm 0.002 \, \textrm{fm.}$ which corresponds to a relative uncertainty of less than $.1\%$ \cite{deVries1987}. As this uncertainty is far smaller than the already neglected NNLO contributions, the parameters in Table \ref{tab:ProtonConst} can be treated as exact for current purposes.

The neutron has no electrical charge and it is correspondingly much more challenging to precisely measure the neutron density distribution. One experimental technique uses measurements from pionic-atoms which allows for indirect determination of the neutron density due to the isospin dependence of the pion-nucleon interaction \cite{Negele1989}. Due to the limitations of this data, the neutron density distribution is usually parameterized in terms of the two-parameter Fermi distribution rather than the model-independent Fourier-Bessel expansion \cite{Garcia1992}. The two-parameter Fermi distribution is given by
\be
\label{eq:2PF}
\rho_{n}\left(r\right) = \frac{\rho_0}{1+e^{\frac{r-c}{z}}}.
\ee
The thickness parameter, $z$, and radial parameter, $c$, describe the shape of the neutron density distribution while $\rho_0$ is a normalization factor. This factor will be chosen such that $\int 4\pi r^{2} \rho\left(r\right)dr = A-Z$.

The neutron thickness parameter, $z$, is usually taken to be equal to the proton thickness parameter for the same nucleus, assuming a two-parameter Fermi distribution for the protons. Coming from the proton distribution, $z$ has a negligible experimental uncertainty but a difficult to quantify systematic uncertainty. Treating this thickness parameter as fixed, it is possible to determine the experimental value and uncertainties of the radial parameter \cite{Garcia1992}. Furthermore, the systematic errors associated with fixing the thickness parameter from the proton distribution can be estimated \cite{Garcia1992}. These systematic errors can be incorporated in the uncertainty in the radial parameter as is done in Table \ref{tab:NeutronConst} of Appendix \ref{app:Nuclear}.

%%%%%%%%%%%%%%%%%%%%%%%%%%%%%%%%%%%%%%%%%%%%%%%%%
\section{Calculation of the Branching Ratio}
\label{sec:BranchingRatio}

The primary quantity of experimental interest is the branching ratio for coherent $\mu-e$ conversion. Expressed in terms of the coherent conversion rate, $\Gamma_{\mu - e}$, and the muon capture rate for the target nucleus, $\omega_{\textrm{capt}}$, the branching ratio is given by
\be
\textrm{BR}(\mu - e)=\frac{\Gamma_{\mu - e}}{\omega_{\textrm{capt}}} .
\ee

To calculate the coherent conversion rate, it will be convenient to treat the CLFV Lagrangians \eqref{eq:scalar-full-eff-Lagrangian} and \eqref{eq:vector-full-eff-Lagrangian} as a series of operators acting along the lepton and nucleon lines of the generic form 
\be
\mathcal{L}_{CLFV} = \frac{1}{\Lambda^{2}} \sum_{j} \bar{e} \mathcal{O}_{L,j} \mu\  \bar{N} \mathcal{O}_{N,j} N .
\ee 

It will be necessary to introduce effective wavefunctions for the nucleons. The isospin index $\alpha$ will be used to distinguish the proton and neutron wavefunctions as $\psi_{\alpha} \left( x \right)$. The wavefunctions will be defined such that $\left| \psi_{\alpha} \left( x \right) \right|^2 = \rho_{\alpha}\left(x\right)$, where $\rho_{\alpha}\left(x\right)$ is the nuclear density distribution as defined in Section \ref{sec:NuclearDensity}. Given these definitions, the wavefunctions are normalized to the nucleon number and not unity. Furthermore, it will be more convenient to work in momentum space and thus one defines the Fourier transformed wavefunctions as 
\be
\widetilde{\psi}_{\alpha} \left( \vec{k}_{N} \right)=\int d^{3}x \; e^{-i \vec{x} \cdot \vec{k}_{N}}\psi_{\alpha} \left( x \right) .
\ee

For the conversion process, the system is initially in a bound state composed of the nucleus and the muon. As the muon is in the ground state, its allowed eigenvalues are $\kappa_{i}=-1$ and $\mu_{i}=\pm\frac{1}{2}$. The eigenvalue of $\kappa_{i}=-1$ is required because the muon's ground state has angular momentum $l=0$. The final state consists of the nucleus and an outgoing electron that may take the eigenvalues $\kappa_{f}=\pm1$ and $\mu_{f}=\pm\frac{1}{2}$. Furthermore, the wavefunction of the electron is also parameterized by the energy of the electron far away from the nuclear potential, $E_{e}$. Neglecting corrections from nuclear recoil which enter at NNLO, conservation of energy requires $E_{e}=m_{\mu}-B_{E}$ where $B_{E}$ is the binding energy of the muon bound state.

The conversion rate can then be expressed as a sum of transition probabilities over all possible spin configurations,
\be
\label{eq:SimpleDecayRate}
\Gamma_{\mu - e} = \frac{1}{2} \displaystyle\sum_{\mu_{i}}\displaystyle\sum_{\mu_{f},\kappa_{f}} \frac{m_{\mu}^5}{\Lambda^4}\left| \tau(E_{e}, \mu_i, \mu_f, \kappa_f ) \right|^{2},
\ee 
where conservation of energy requires $E_{e}=m_{\mu}-B_{E}$. The conversion amplitude may be written in a dimensionless form as,
\be
\begin{aligned}
\label{eq:SimpleDecayAmplitude}
\tau(E_{e}, \mu_i, \mu_f, \kappa_f ) = \frac{1}{m_{\mu}^{5/2}}\sum_{j} \int \frac{d^{3}k_{N}'}{{\left(2\pi\right)^3}} \int \frac{d^{3}k_{N}}{{\left(2\pi\right)^3}} & \left [\int d^{3}x e^{i\left(\vec{k}_{N}-\vec{k}_{N}'\right)\cdot\vec{x}} \psi^{\left(e\right)\dagger}_{\kappa_{f},\mu_{f},E_{e}}\left(x\right) \mathcal{O}_{L,j} \psi^{\left(\mu\right)}_{-1,\mu_{i}}\left(x\right) \right] \\
& \cdot \left[ \widetilde{\psi}^{*}_{\alpha'}\left(\vec{k}_{N}'\right) \mathcal{O}_{N,j}^{\alpha,\alpha'} \left( \left| \vec{k}_{N}'- \vec{k}_{N} \right| \right) \widetilde{\psi}_{\alpha} \left( \vec{k}_{N} \right) \right].
\end{aligned}
\ee
The isospin indices $\alpha$ and $\alpha'$ have been introduced for the hadronic operator as it may have isospin dependence, as occurs in the case of vector-mediated conversion. The summation over the index $j$ corresponds to summing over the contributions of each operator in the CLFV Lagrangian.

The structure of the phase space integrals in \eqref{eq:SimpleDecayAmplitude} does not depend on the model of CLFV and thus it is straightfoward to numerically evaluate these overlap integrals for each possible operator in the Lagrangians \eqref{eq:scalar-full-eff-Lagrangian} and \eqref{eq:vector-full-eff-Lagrangian}. This procedure is detailed in Appendix \ref{app:Overlap}, and the numerical values and accompanying uncertainties for the phase space integrals of ${}^{27}$Al are given in Table \ref{tab:Overlap} of the same appendix.

As stated previously, there are eight possible spin configurations. However, there is a two-fold symmetry in the choice of overall sign for the spins. This reduces the number of independent configurations to only four. For compactness of notation, an index $w\in\left\{1,2,3,4\right\}$ will be used to denote each unique configuration. The relationship between all possible spin configurations and $w$ is given in Table \ref{tab:SpinConfig} of Appendix \ref{app:Overlap}. The branching ratio can then be written in terms of four separate amplitudes, one for each configuration, leading to Eq.~(\ref{eq:brmaster}). The corresponding expression for the $\tau_S^{(w)}$ and $\tau_V^{(w)}$ are given in Eqs.~(\ref{eq:ConvAmpl-Scalar}) and (\ref{eq:ConvAmpl-Vector}), respectively. The expressions Eqs.~(\ref{eq:brmaster},\ref{eq:ConvAmpl-Scalar},\ref{eq:ConvAmpl-Vector}) and the model independent parameters of Tables \ref{tab:BR-Scalar} and \ref{tab:BR-Vector} allow one to start with an arbitrary model of CLFV and calculate in a straightforward manner the coherent conversion branching ratio including NLO contributions and uncertainties. These expressions and their model independent parameters constitute the primary results of this paper and their use is summarized in Appendix \ref{app:BR-Formula}.

%%%%%%%%%%%%%%%%%%%%%%%%%%%%%%%%%%%%%%%%%%%%%%%%%
\section{Discussion and Analysis}\label{sec:Discussion}

Having expressed $\textrm{BR}(\mu - e)$ in terms of products of CLFV model-dependent Wilson coefficients and model-independent SM factors, we now discuss the implications in terms of sensitivity to various CLFV scenarios. We first consider the case of scalar-mediated conversion. The model-independent parameter $\alpha_{S,ud}^{(w)}$ is given by
\be
\begin{multlined}[t][.87\textwidth]
\alpha_{S,ud}^{(w)} = \sqrt{ \frac{m_{\mu}}{\omega_{\textrm{capt}}} } \left( \frac{m_{\mu}}{4 \pi v} \right)^2 \left[ \frac{\sigma_{\pi N}}{2 \hat{m}} \left( I^{(w)}_{S,p} + I^{(w)}_{S,n} \right) - \frac{3B_{0}K_{F} \mathring{g}_{A}^{2}}{64\pi \mathring{f}_{\pi}^{2}} f^{SI}_{\textrm{eff}}\left( I^{(w)}_{S,p} + I^{(w)}_{S,n} \right) \right. \\
\left. - \frac{3B_{0}m_{\pi} \mathring{g}_{A}^{2}}{64\pi \mathring{f}_{\pi}^{2}} \Delta_S^{(w)} \right] ,
\end{multlined}
\ee
where
\be
\Delta_S^{(w)}= \left({\tilde I}_{S,p}^{(w)}+{\tilde I}_{S,n}^{(w)}\right)-\left(I_{S,p}^{(w)}+I_{S,n}^{(w)}\right).
\label{eq:definition-delta}
\ee

As is done in Table \ref{tab:BR-Scalar}, one can consider the LO, NLO loop, and NLO two-nucleon contributions independently. Consider the ratio of the NLO loop contribution to the LO contribution,
\be
\frac{-\alpha_{S,ud}^{(w)}(\mathrm{NLO\ loop})}{\alpha_{S,ud}^{(w)}(\mathrm{LO})} = \frac{ \left( \frac{3 B_0m_\pi \mathring{g}_A^2}{64\pi \mathring{f}_\pi^2}\right) \ \Delta_S^{(w)} }{ \frac{\sigma_{\pi N}}{{2\hat m}}\left(I_{S,p}^{(w)}+I_{S,n}^{(w)}\right)} = \left[ \frac{2\hat m}{\sigma_{\pi N}} \left( \frac{3 B_0m_\pi \mathring{g}_A^2}{64\pi \mathring{f}_\pi^2}\right) \right] \cdot \left[ \frac{ \Delta_S^{(w)} }{ \left(I_{S,p}^{(w)}+I_{S,n}^{(w)}\right)} \right] .
\ee
We have written the ratio as the product of two terms. The first term only depends on the dimensionful low-energy constants parameterizing the relative strength of the LO and NLO couplings. The second term is kinematic in nature, arising from overlap integrals, and is dependent on the spin configuration. Numerically, one finds
\begin{align}
\frac{2\hat m}{\sigma_{\pi N}} \left( \frac{3 B_0m_\pi \mathring{g}_A^2}{64\pi \mathring{f}_\pi^2}\right) & = 0.160 \pm 0.029 \; , \label{eq:coupling-ratio-LO-NLO} \\
\frac{ \Delta_S^{(w)} }{ \left(I_{S,p}^{(w)}+I_{S,n}^{(w)}\right)} & = 
\begin{cases}
0.261 \pm 0.026 & w = 1 \\
0.260 \pm 0.027 & w = 2 \\
0.260 \pm 0.027 & w = 3 \\
0.259 \pm 0.025 & w = 4 
\end{cases} \; . \label{eq:kinematic-ratio-LO-NLO}
\end{align}

As can be seen from \eqref{eq:coupling-ratio-LO-NLO}, the NLO contribution is small compared to the LO contribution just due to the hierarchy of their dimensionful parameters, exactly as expected from ChPT. However, \eqref{eq:kinematic-ratio-LO-NLO} shows that the NLO term is additionally suppressed by kinematic considerations. As discussed in Section \ref{sec:Introduction}, the NLO loop contribution depends on $\Delta_S^{(w)}$ which vanishes in the limit of zero momentum transfer. Due to the relatively low momentum transfer involved in coherent conversion, $\left| q_T  \right| \approx m_{\mu}$, this further reduce the size of the NLO loop contribution. Taken together, \eqref{eq:coupling-ratio-LO-NLO} and \eqref{eq:kinematic-ratio-LO-NLO} result in the NLO loop contribution being particularly small --  roughly $5\%$ of the LO contribution. The NLO loop contribution is sufficiently small that even the parametric uncertainty in the LO contribution is larger than it.

This should be contrasted with the NLO  two-body contribution, which is sizable and may appreciably reduce the conversion amplitude. The ratio of the NLO two-nucleon contribution to the LO contribution is
\be
\frac{-\alpha_{S,ud}^{(w)}(\mathrm{NLO\ NN})}{\alpha_{S,ud}^{(w)}(\mathrm{LO})} = \frac{ \frac{3B_{0}K_{F} \mathring{g}_{A}^{2}}{64\pi \mathring{f}_{\pi}^{2}} f^{SI}_{\textrm{eff}}\left( I^{(w)}_{S,p} + I^{(w)}_{S,n} \right) }{ \frac{\sigma_{\pi N}}{2 \hat{m}} \left( I^{(w)}_{S,p} + I^{(w)}_{S,n} \right) } = \frac{ 2 \hat{m} }{ \sigma_{\pi N} } \left( \frac{3B_{0}K_{F} \mathring{g}_{A}^{2}}{64\pi \mathring{f}_{\pi}^{2}} \right) f^{SI}_{\textrm{eff}} = 0.29^{+0.06}_{-0.16} \; .
\ee
While there is significant uncertainty in the value of the two-nucleon contribution due to the one-body averaging procedure of Section \ref{sec:Effective-One-Nucleon}, the two-nucleon contribution is expected to be $15\%-30\%$ of the LO contribution. As the two-nucleon contribution has the opposite sign of the LO contribution, this can result in the coherent conversion branching ratio decreasing by as much as $25\%-50\%$. It may seem surprising that the NLO two-nucleon contribution is so much larger than the loop contribution but this difference is due to the fact that the loop contribution is suppressed for kinematic reasons encapsulated in $\Delta_S^{(w)}$ which are unrelated to the chiral expansion of ChPT. Given the potentially significant impact of the NLO two-nucleon contribution on the sensitivity of $\textrm{BR}(\mu - e)$ to scalar-mediated interactions,  a state-of-the-art many-body computation of this contribution should be performed.

Lastly, we consider the relative size of parametric uncertainties in scalar-mediated conversion to the theoretical uncertainties which arise from our neglect of NNLO contributions. For the LO contribution, the dominant uncertainty is in determining the quark content of the nucleons. Ignoring factors common to all the model-independent parameters, $\alpha_{S,ud}^{(w)}(\mathrm{LO}) = \frac{\sigma_{\pi N}}{2 \hat{m}} \left( I^{(w)}_{S,p} + I^{(w)}_{S,n} \right)$. Both the isospin average quark mass and the sum of overlap integrals are known to within $\sim2\%$, see Tables \ref{tab:Constants} and \ref{tab:Overlap} of Appendices \ref{app:Constants} and \ref{app:Overlap} respectively. However, the pion-nucleon sigma-term, $\sigma_{\pi N}$, has a relative uncertainty of $\sim17\%$, see Table \ref{tab:Constants}. This is significantly larger than the NLO loop contribution and is comparable in size to the NLO two-nucleon contribution. Even if the NNLO contributions are comparable in size to the NLO loop contribution and are $\sim5\%$ of the LO term, significant improvements must be made in the determination of the pion-nucleon sigma-term before the theoretical uncertainty from neglecting NNLO corrections becomes relevant.

We now turn our attention to the case of vector-mediated coherent conversion. As has been shown in Section \ref{sec:Vector-Mediated-Conversion}, the NLO contributions to the vector-mediated process are spin-dependent and suppressed by a factor of $1/A$. This suppression makes them comparable in size to the already neglected NNLO contributions. Consequently, the model-independent parameters are completely determined by the leading-order contributions
\begin{align}
&\alpha_{V,u}^{(w)} = \sqrt{ \frac{m_{\mu}}{\omega_{\textrm{capt}}} } \left( \frac{m_{\mu}}{4 \pi v} \right)^2 \left( 2 I^{(w)}_{V,p} + I^{(w)}_{V,n} \right) , \label{eq:a-VectorUp} \\
&\alpha_{V,d}^{(w)} = \sqrt{ \frac{m_{\mu}}{\omega_{\textrm{capt}}} } \left( \frac{m_{\mu}}{4 \pi v} \right)^2 \left( I^{(w)}_{V,p} + 2 I^{(w)}_{V,n} \right) . \label{eq:a-VectorDown}
\end{align}
These parameters are known to within $\sim2\%$ and the dominant uncertainty is from the overlap integrals, which in turn is a reflection of uncertainties in the neutron distribution of ${}^{27}\textrm{Al}$, see Table \ref{tab:NeutronConst} of Appendix \ref{app:Nuclear}. Of course, these are parametric uncertainties and theoretical uncertainties from the neglect of NNLO terms are not included. Given that the NLO contributions were suppressed, it is difficult to estimate the magnitude of the NNLO contributions. However, one naively expects NNLO corrections in $SU(2)$ HBChPT to contribute at roughly the two percent level and the NLO loop correction for scalar mediated conversion was found to be five percent of the LO term. Thus, one may conservatively estimate the theoretical uncertainty to be roughly five percent.

%%%%%%%%%%%%%%%%%%%%%%%%%%%%%%%%%%%%%%%%%%%%%%%%%
\section{Conclusions}\label{sec:Conclusion}

In this work, we have performed an analysis of coherent $\mu-e$ conversion at next-to-leading order and have carefully tracked possible sources of uncertainty. The primary results of this analysis are the expressions Eqs.~(\ref{eq:brmaster},\ref{eq:ConvAmpl-Scalar},\ref{eq:ConvAmpl-Vector}) and the corresponding model independent parameters of Tables \ref{tab:BR-Scalar} and \ref{tab:BR-Vector}. These results are summarized in Appendix \ref{app:BR-Formula}.

Starting with a CLFV Lagrangian of the generic form \eqref{eq:CLFV-Lagrangian-generic}, one may define the Wilson coefficients \eqref{eq:eff-lep-coef-scalar}-\eqref{eq:eff-lep-coef-mass}. It is then straightforward to use Eqs.~(\ref{eq:brmaster},\ref{eq:ConvAmpl-Scalar},\ref{eq:ConvAmpl-Vector}) and the corresponding model-independent parameters to calculate the branching ratio for coherent conversion at next-to-leading order including uncertainties. Similarly, one can use these expressions to determine the permitted regions of parameter space in the event of a detection or non-detection at the upcoming \noun{Mu2E} and \noun{COMET} experiments.

In our analysis of scalar-mediated CLFV, we find that the contributions from the next-to-leading order loop diagram are generally small. However, the contributions from the next-to-leading order two-body diagram have the opposite sign of the leading order contribution and could be up to $30\%$ of its size. This can result in an order one change in the branching ratio for a model of CLFV. For a fixed mediator mass, the sensitivity of the upcoming \noun{Mu2E} and \noun{COMET} experiments can be reduced by up to a factor of two.

In the case of scalar-mediated conversion, we find that the dominant source of uncertainty is the determination of the nucleon sigma-terms and quark masses. These uncertainties result in a $30\%$ uncertainty in the amplitude for coherent conversion. This severely limits the ability of a single target detector to discriminate different models of CLFV. Generally, these hadronic uncertainties need to be improved by at least a factor of four before NNLO corrections become relevant. Another significant source of uncertainty comes from the one-body averaging of the two-nucleon effective operator. A more careful treatment of this operator including a full many-body treatment of the nucleus would result in improved uncertainties.

Compared to scalar-mediated conversion, vector-mediated conversion has significantly smaller uncertainties. The dominant source of uncertainty comes from the determination of the neutron distribution in ${}^{27}\textrm{Al}$ and this only contributes at the two percent level. This is comparable to the theoretical uncertainties from the neglected NNLO corrections. As such, to improve the precision of the vector-mediated case it will be necessary to calculate the NNLO contributions. This will be technically challenging as it requires a careful treatment of the many-body nuclear wavefunction with spin-dependence.

While the analysis presented here is specific to ${}^{27}\textrm{Al}$, it should be straightforward to extend the present approach to other potential targets. As has been shown in the literature \cite{Cirigliano2009}, multiple targets will be required in the event of detection to determine the channel of CLFV. Given the large hadronic uncertainties in the branching ratio for scalar-mediated conversion, the use of multiple targets is highly desirable because it should allow an improved determination of CLFV model parameters over what is naively indicated by the hadronic uncertainties.

%%%%%%%%%%%%%%%%%%%%%%%%%%%%%%%%%%%%%%%%%%%%%%%%
\section*{Acknowledgements}

We thank Mark B. Wise for many helpful discussions throughout the course of this work. This work was supported in part under U.S. Department of Energy contracts DE-SC0011632 (AB) and DE-SC0011095 (MJRM).

\newpage

%%%%%%%%%%%%%%%%%%%%%%%%%%%%%%%%%%%%%%%%%%%%%%%%
\begin{appendix}

\section{Chiral Lagrangian}
\label{app:ChiralLagrangian}

Ignoring the stress-energy tensor, the quark-level CLFV Lagrangian, \eqref{eq:CLFV-Lagrangian-quarks}, written in terms of the CLFV currents, \eqref{eq:scalar-CLFV-current} and \eqref{eq:vector-CLFV-current}, is given by
\be
\mathcal{L}_{\mathrm{CLFV}} = \displaystyle\sum_{f=\mathrm{u,d,s}} \frac{1}{\Lambda^{2}} J_{f} \bar{q}_{f}q_{f} + \displaystyle\sum_{f=\mathrm{u,d,s}}\frac{1}{\Lambda^{2}} J_{f}^{\nu} \bar{q}_{f}\gamma_{\nu}q_{f}. \label{eq:quark-level-CLFV-current-Lagrangian}
\ee
HBChPT can then be used to relate \eqref{eq:quark-level-CLFV-current-Lagrangian} to the physics of nucleons and mesons. The resulting effective theory will have several unknown LECs that can be determined by matching onto experimental determinations of hadronic matrix elements. Through the electromagnetic interaction, the matrix elements for the vector current $\left\langle N \right| \bar{q}_{f}\gamma_{\nu}q_{f} \left| N \right\rangle$ are known in terms of the Pauli and Dirac or Sachs form factors. For scalar quark currents, the relevant matrix elements are $\left\langle N \right| m_f \bar{q}_{f} q_{f} \left| N \right\rangle$, not $\left\langle N \right| \bar{q}_{f} q_{f} \left| N \right\rangle$. To make contact with the known matrix elements, we introduce factors of the quark mass to rewrite the scalar CLFV term of \eqref{eq:quark-level-CLFV-current-Lagrangian} as
\be
\mathcal{L}_{\mathrm{CLFV}} = \displaystyle\sum_{f=\mathrm{u,d,s}} \left( \frac{J_{f}}{m_f \Lambda^{2}} \right) m_f \bar{q}_{f}q_{f}.
\ee
This has the same form as the operator responsible for insertions of the quark mass. Explicitly including this term in the Lagrangian,
\be
\mathcal{L} = \displaystyle\sum_{f=\mathrm{u,d,s}} \left[ -1 + \left( \frac{J_{f}}{m_f \Lambda^{2}} \right) \right] m_f \bar{q}_{f}q_{f} \label{eq:factorized-CLFV-Lagrangian}
\ee
Given the form of \eqref{eq:factorized-CLFV-Lagrangian}, it is apparent that the scalar CLFV current enters the chiral Lagrangian with the same matrix elements as the quark mass insertion. However, the scalar CLFV current also carries inverse factors of $\Lambda^2$ and $m_f$. Thus, up to these additional factors, the LECs of the effective theory can be expressed in terms of known nuclear matrix elements.

In constructing the Lagrangian for HBChPT, one has dynamical fields corresponding to the pions ($\pi^{0}$, $\pi^{\pm}$) and nucleons ($\Psi_{P}$, $\Psi_{N}$) along with insertions of the CLFV currents. These currents and dynamical fields can be organized into a collection of objects with well defined transformation properties under the chiral $SU\left(2\right)$ symmetry,
\begin{table}[ht]
\centering
\begin{tabular}{p{0.3\textwidth} p{0.4\textwidth}}
$\phi = \left(\begin{array}{cc}
\pi^{0} & \sqrt{2}\pi^{+}\\
\sqrt{2}\pi^{-} & -\pi^{0}
\end{array}\right)$ & $v^{\mu} = \frac{1}{\Lambda^{2}} \cdot \frac{J^{\mu}_{u}-J^{\mu}_{d}}{2}\left(\begin{array}{cc}
1 & 0\\
0 & -1
\end{array}\right)$ \\
$U=\exp\left(\frac{i\phi}{\mathring{f}_{\pi}}\right)$ &  $v^{\left(s\right)\mu} = \frac{1}{\Lambda^{2}} \cdot \frac{3}{2}\left(J^{\mu}_{u}+J^{\mu}_{d}\right)$ \\
$u=\exp\left(\frac{i\phi}{2\mathring{f}_{\pi}}\right)$ & $v_{s}^{\left(s\right)\mu} = \frac{1}{\Lambda^{2}} J_{s}^{\mu}$ \\
$N=\left(\begin{array}{c}
\Psi_{P}\\
\Psi_{N}
\end{array}\right)$ & $u_{\mu}=i\left[u^{\dagger}\left(\partial_{\mu}-iv_{\mu}\right)u-u\left(\partial_{\mu}-iv_{\mu}\right)u^{\dagger}\right]$ \\
$\chi=-2B_{0}\frac{1}{\Lambda^{2}} \left(\begin{array}{cc}
J_{u} & 0\\
0 & J_{d}
\end{array}\right)$ & $\chi_{s}=-2B_{0}\frac{1}{\Lambda^{2}}J_{s}$\\
\end{tabular}
\end{table} 

In these expressions, $\mathring{f}_{\pi}$ is the tree-level pion decay constant in the chiral limit and $B_0$ normalizes the scalar sources. The chiral Lagrangian can then be constructed from these objects by considering all possible combinations that are invariant under chiral $SU\left(2\right)$ transformations. These terms can be grouped by chiral order so that the chiral Lagrangian corresponds to a well defined expansion in chiral powers. In our power counting, we will assign the CLFV currents $J_f$ and $J_f^{\nu}$ chiral order $\mathcal{O} \left( 1 \right)$ as they do not scale with the quark mass.

As complete expressions for the chiral Lagrangian beyond NLO can be found in the literature \cite{Ecker1996, Ecker1997, Bernard1996, Meissner1998}, only terms that include CLFV operators will be listed here. The relevant CLFV terms present in the pionic Lagrangians are given by,
\be
\mathcal{L}_{\pi\pi}^{(0)} = \frac{\mathring{f}_{\pi}^{2}}{4} Tr \left[ \chi \left(U^{\dagger} + U\right) \right],
\ee
\be
\mathcal{L}_{\pi\pi}^{(1)} = \frac{\mathring{f}_{\pi}^{2}}{2} Tr\left[i\left(\partial_{\mu}U^{\dagger}U + \partial_{\mu}UU^{\dagger}\right)v^{\mu}\right].
\ee

Fixing a reference velocity $V^{\mu}$ for HBChPT, the CLFV terms in the pion-nucleon Lagrangians are,
\begin{multline}
\mathcal{L}_{\pi N}^{(0)} = \bar{N} \left[ \frac{1}{2} V^{\mu}\left( u^{\dagger}v_{\mu}u + uv_{\mu}u^{\dagger} + 2v_{\mu}^{(s)} \right) + d_{1}^{S}\chi_{s} \right.\\
 \left. + \bar{c}_{5} \left(\chi\left(U + U^{\dagger}\right) - \frac{1}{2} Tr\left[ \chi\left(U + U^{\dagger} \right)\right]\right) + \bar{c}_{1} \, Tr\left[ \chi\left( U + U^{\dagger} \right) \right]  \right]N,
\end{multline}

\be
\begin{aligned}
\mathcal{L}_{\pi N}^{(1)} =
\bar{N} & \left[ -i\frac{1}{2 M_{N}} V^{\mu}V^{\nu}  \left( \partial_{\mu} v_{\nu} + 2 v_{\nu} \partial_{\mu} + \partial_{\mu}v_{\nu}^{(s)} + 2v_{\nu}^{(s)}\partial_{\mu} \right) \right. \\
&+ i\frac{1}{2 M_{N}} \left( \partial_{\mu} v^{\mu} +2 v_{\mu} \partial^{\mu} + \partial^{\mu}v_{\mu}^{(s)} + 2v_{\mu}^{(s)}\partial^{\mu} \right)\\
&-i\frac{1}{2 M_{N}} \left[ S^{\mu} , S^{\nu} \right] \left( 1 + \mathring{k}_{V} \right) \left( \partial_{\mu}v_{\nu} - \partial_{\nu}v_{\mu} \right) \\
&-i\frac{1}{2M_{N}} \left[ S^{\mu} , S^{\nu} \right] \left( 1 + \mathring{k_{s}} \right) \left( \partial_{\mu}v_{\nu}^{(s)} - \partial_{\nu}v_{\mu}^{(s)} \right)\\
&\left. -i\frac{1}{2M_{N}} \left[ S^{\mu} , S^{\nu} \right] d_{1}^{V} \left( \partial_{\mu}v_{s\,\nu}^{(s)} - \partial_{\nu}v_{s\,\mu}^{(s)} \right) \right]N ,
\end{aligned}
\ee
where $S^{\mu}$ is the spin operator for HBChPT. In these expressions for the pion-nucleon Lagrangian, the coefficients $d_{1}^{V}$ and $d_{1}^{S}$ have been introduced. These are new LECs that correspond to strange quark operators that do not normally appear in $SU(2)$ HBChPT. The coefficients $\bar{c}_{1}$ and $\bar{c}_{5}$ have also been introduced and should be distinguished from the usual LECs $c_{1}$ and $c_{5}$ of $SU \left( 2 \right)$ HBChPT. As explained previously, the LECs of the CLFV effective theory differ from the usual matrix elements by a factor of $1 / m_q$, see, {\em e.g.}, \eqref{eq:LEC-to-matrix-element}.

An additional set of terms of the form $\bar{N} \left( Tr \left[ \chi_{+} \right] iV \cdot \partial + \ldots \right) N$ should also appear in $\mathcal{L}_{\pi N}^{(2)}$. However, for coherent $\mu-e$ conversion, these operators will only appear as insertions on an on-shell nucleon line. For on-shell momenta, $V \cdot k$ is of order  $\mathcal{O}\left(q^{2}\right)$ and thus these operators should be treated as $\mathcal{O}\left(q^{3}\right)$. Thus, these terms can be neglected.

It is worth noting that $\left[ S^{\mu} , S^{\nu} \right] \left( \partial_{\mu}v_{s\,\nu}^{(s)} - \partial_{\nu}v_{s\,\mu}^{(s)} \right)$ is not the only $SU\left(2\right)$ invariant term one could write for a generic isoscalar operator. Naively one could write additional terms involving $v_{s\,\mu}^{(s)}$, however, in addition to being isoscalar $v_{s\,\mu}^{(s)}$ carries strangeness. As the nucleons do not carry net strangeness, the only allowable term at this order is $\left[ S^{\mu} , S^{\nu} \right] \left( \partial_{\mu}v_{s\,\nu}^{(s)} - \partial_{\nu}v_{s\,\mu}^{(s)} \right)$.

\newpage
\section{Low-Energy Constants of the Isoscalar Strange Operators}
\label{sec:LECs-and-LQCD}

The low-energy constants that appear in the effective Lagrangian of Appendix \ref{app:ChiralLagrangian} can be assigned numerical values by making contact with experimental results. This is done by matching analytical expressions for nucleon matrix elements in HBChPT onto those from QCD. However, the normalization schemes for nucleon states in QCD and HBChPT are different. HBChPT treats nucleons as non-relativistic fields with the appropriate non-relativistic normalization while QCD is fully relativistic. The differences between these schemes are of order $\mathcal{O}\left( \frac{k^2}{2 m_N}\right)$ and thus only enter at NNLO. As such, these schemes may be treated as equivalent for present purposes.

Most of the LECs appear in standard $SU(2)$ HBChPT and are well known, however, the additional constants introduced by including the isoscalar strange operators must be determined. For the scalar strange operator, there is only one unknown LEC. Comparing terms in the chiral and QCD Lagrangians, there is the equivalency
\be
-m_s \bar{q_{s}}q_{s} \simeq \bar{N} \left[ 2B_{0}d_{1}^{S} m_s + \mathcal{O} \left( q^2 \right) \right] N.
\ee
Using the matrix element for the contribution of the strange quark condensate to the nucleon mass, $m_N f_{s}^{N}=\left\langle N(0) \right| m_{s}\bar{q_{s}}q_{s} \left| N(0) \right\rangle$, one immediately arrives at the result
\be
\frac{m_{N} f_{s}^{N}}{m_{s}}  = -2B_{0} d_{1}^{S}.
\ee

For the vector strange operator, one must compare the vector current to the electric and magnetic nucleon form factors. Written in terms of the Sachs form factors and only keeping terms through NLO \cite{Perdrisat2007},
\be
\left \langle N \left( k' \right) \left| \bar{q}_{s}\gamma_{\mu}q_{s} \right|N \left( k \right) \right \rangle = \overline{u}\left( k' \right) \left[ \gamma_{\mu} G_{E}^{S} \left( q_{T}^{2} \right) + \frac{i\sigma_{\mu\nu}q_{T}^{\nu}}{2m_{N}} \left( G_{M}^{S} \left( q_{T}^{2} \right) - G_{E}^{S} \left( q_{T}^{2} \right) \right) \right] u\left( k \right).
\ee
These form factors are functions of the three-momentum transfer, $q_{T}^{2}$. As the momentum transfer is much smaller than the nucleon mass, the Sachs form factors can be rewritten as series expansions in the momentum transfer. Because nucleons have no net strangeness, the leading order terms of these expansions are $G_{E}^{S}\left(q_{T}^{2}\right) = \rho_{s}\frac{q_{T}^{2}}{4M_{N}}$ and $G_{M}^{S}\left(q_{T}^{2}\right) = \mu_{s} + \mathcal{O}\left(q_{T}^{2}\right)$ where $\rho_{s}$ is the strangeness radius and $\mu_{s}$ is the strange magnetic moment \cite{Pate2014}. Thus, keeping terms only through NLO the matrix element is given by
\be
\left \langle N \left( k' \right) \left| \bar{q}_{s}\gamma_{\mu}q_{s} \right|N \left( k \right) \right \rangle = \overline{u}\left(k'\right) \left[ -\frac{1}{M_{N}} \left[ S_{\mu} , S_{\nu} \right] q_{T}^{\nu}\mu_{s} \right] u\left(k\right).
\ee
This matrix element must then be matched onto the corresponding matrix element for the strange vector current in HBChPT. This vector current can be directly read off of the Lagrangian \eqref{eq:vector-CLFV-momentum}. This fixes the value of the unknown LEC to be $d_{1}^{V}=\mu_{s}$.

\newpage
\section{Values of Low-Energy Constants and Physical Quantities}
\label{app:Constants}

All parameters that depend on renormalization are given in the $\overline{MS}$ scheme at $\mu=2\,\textrm{GeV}$ except where otherwise noted. All values taken from the world lattice data \cite{FLAG2016} make use of results from $N_f = 2 + 1$ simulations whenever possible.

\begin{table}[h]
\centering
\begin{tabular}{l|r|l|l}
Quantity & Accepted Value & Source & Notes\\ \hline
$\hat{m}$ & $3.373\pm 0.080 \,\textrm{MeV}$ & \cite{FLAG2016, Durr2011, Durr2011_2, McNeile2010, Blum2016, Bazavov:2010yq} & \\
$m_{u} / m_{d}$ & $0.46\pm0.03 \,\textrm{MeV}$ & \cite{FLAG2016, Durr2011, Durr2011_2, McNeile2010, Blum2016, Bazavov:2010yq} & \\
$m_{s}$ & $92.0\pm 2.1 \,\textrm{MeV}$ & \cite{FLAG2016, Bazavov:2009fk, Durr2011,Durr2011_2, McNeile2010, Blum2016} & \\
$f_{\pi}$ & $92.07\pm0.99 \,\textrm{MeV}$ & \cite{FLAG2016, Follana2008, Bazavov:2010hj, Arthur2013} &  See Footnote \footnote[1]{Due to a difference in definitions, the value of $f_\pi$ presented here is the value from \cite{FLAG2016} divided by $\sqrt{2}$.}\\
$f_{\pi} / \mathring{f}_{\pi}$ & $1.064\pm0.007$ & \cite{FLAG2016, Blum2016, Bazavov:2010hj, Borsanyi2013, Durr2014, Beane2012}\\
$\Sigma$ & $274\pm3 \,\textrm{MeV}$ & \cite{FLAG2016, Blum2016, Bazavov:2010yq, Borsanyi2013, Durr2014}\\
$g_{A}$ & $1.2759\pm0.0045$ & \cite{UCNA2010}\\
$\sigma_{\pi N}$ & $52 \pm 9 \,\textrm{MeV}$ & \cite{Alvarez-Ruso:2014}\\
$f_{s}$ & $0.043\pm0.011$ & \cite{Junnarkar2013}\\
$v$ & $246.220 \,\textrm{GeV}$ & \cite{PDG2016} & See Footnote \footnote[2]{These quantities presented in \cite{PDG2016} are known to a precision far beyond the other values in this table. As such, they are presented without uncertainties.} \\
$m_{\pi}$ & $138.039 \,\textrm{MeV}$ & \cite{PDG2016} & Isospin averaged pole mass \footnotemark[2] \\
$m_{N}$ & $938.919 \,\textrm{MeV}$ & \cite{PDG2016} & Isospin averaged pole mass \footnotemark[2] \\
$m_{\mu}$ & $105.658 \,\textrm{MeV}$ & \cite{PDG2016} & Pole mass \footnotemark[2] \\
$K_F$ & $238\pm5 \, \textrm{MeV}$ & \cite{Moniz1971} & For ${}^{27}_{13}\textrm{Al}$ \footnote[3]{From linear interpolation between the experimentally measured Fermi momenta of ${}_{12}^{24}\textrm{Mg}$ and ${}_{20}^{40}\textrm{Ca}$} \\
$\omega_{\textrm{capt}}$ & $705.4 \pm 1.3 \, \textrm{ms}^{-1}$ & \cite{Suzuki1987} & For ${}^{27}_{13}\textrm{Al}$ \\
 \hline
 $\omega_{\textrm{capt}}$ & $464.30 \pm 0.86 \, \textrm{peV}$ & Derived & Unit conversion with $\hbar = 1$\\
$\xi$ & $0.37\pm 0.02$ & Derived & $\xi\equiv\frac{1 - m_u/m_d}{1 + m_u/m_d}$\\
$\mathring{f}_{\pi}$ & $86.5 \pm 1.1 \,\textrm{MeV}$ & Derived &  $\mathring{f}_{\pi} \equiv f_\pi \left( \frac{f_{\pi}}{\mathring{f}_{\pi}}\right)^{-1}$\\
$B_{0}$ & $2.75\pm0.11 \,\textrm{GeV}$ & Derived & $B_{0} = \frac{\Sigma^3}{\mathring{f}_{\pi}^{2}}$\\
$\sigma_{s N}$ & $40\pm10 \,\textrm{MeV}$ & Derived & $\sigma_{s N} \equiv f_s m_N$\\
\end{tabular}
\caption{Table of low-energy constants\label{tab:Constants}}
\end{table}

\newpage
\section{Momentum Dependence of Effective One-Body Operator}
\label{app:OneBody}

In Section \ref{sec:Scalar-Mediated-Conversion}, it was shown that a two-nucleon operator enters the effective scalar CLFV Lagrangian at NLO. This two-nucleon operator can be reduced to an effective one-nucleon operator,  \eqref{eq:eff-one-nucleon-momentum}, by treating the nuclear core as a degenerate Fermi gas and averaging over core nucleons as was explained in Section \ref{sec:Effective-One-Nucleon}. The dependence of \eqref{eq:eff-one-nucleon-momentum} on both the momentum transfered to the nucleus, $q_T$, and the average of the initial and final nucleon momenta, $k$, is encapsulated in the dimensionless functions $f^{SI}$ and $f^{SD}$. For compactness of notation, define the following dimensionless parameters in terms of the Fermi momentum $K_F$,
\be
\overline{q} = \frac{q_T}{K_F}, \quad
\overline{k} = \frac{k}{K_F}, \quad
\overline{m} = \frac{m_\pi}{K_F}.
\ee
In terms of these dimensionless parameters, the functions $f^{SI}$ and $f^{SD}$ can be expressed as,
\be
\begin{aligned}
f^{SI}\left(\overline{q}, \overline{k}\right) = \frac{1}{2\pi} \int_{-1/2}^{1/2} & d\beta \left[ 2 \left( 1 + \frac{ -\beta\, \overline{k} \cdot \overline{q} + \beta^2\, \overline{q}^2 }{ \overline{k}^2 - 2 \beta\, \overline{k} \cdot \overline{q} + \beta^2\, \overline{q}^2 } \right) \right. \\
& - \left( \frac{ 4 \left( \frac{1}{4} - \beta^2 \right) \overline{q}^2 + 3 \overline{m}^2 }{ \sqrt{ \left( \frac{1}{4} - \beta^2 \right) \overline{q}^2 + \overline{m}^2 } } \right) \arctan \left( \frac{ 2\sqrt{ \overline{m}^2 + \left( \frac{1}{4} - \beta^2 \right) \overline{q}^2 } }{ \overline{m}^2 + \frac{1}{4} \overline{q}^2 - 2 \beta\, \overline{k} \cdot \overline{q} + \overline{k}^2 - 1} \right) \\
& + \frac{ 1 }{ 2 \sqrt{ \overline{k}^2 - 2 \beta\, \overline{k} \cdot \overline{q} + \beta^2\, \overline{q}^2 } } \cdot \left( 1 + 2 \overline{m}^2 + \left( \frac{3}{4} - 4 \beta^2 \right) \overline{q}^2 - \overline{k}^2 \right. \\
& \left. + 2 \beta\, \overline{k} \cdot \overline{q} +  \beta \frac{ \left( 1 + \frac{1}{4} \overline{q}^2 + \overline{m}^2 + \overline{k}^2 - 2\beta\, \overline{k} \cdot \overline{q} \right) \left( \overline{k} \cdot \overline{q} - \beta \, \overline{q}^2 \right) }{ \overline{k}^2 - 2 \beta\, \overline{k} \cdot \overline{q} + \overline{q}^2 } \right) \\
& \left. \cdot\ln\left( \frac{1+2\sqrt{ \overline{k}^2 - 2\beta\, \overline{k} \cdot \overline{q} + \beta^2\, \overline{q}^2 } + \overline{k}^2 - 2\beta\, \overline{k} \cdot \overline{q} + \frac{1}{4} \overline{q}^2 + \overline{m}^2 }{1 - 2 \sqrt{ \overline{k}^2 - 2\beta\, \overline{k} \cdot \overline{q} + \beta^2\, \overline{q}^2 } + \overline{k}^2 - 2\beta\, \overline{k} \cdot \overline{q} + \frac{1}{4} \overline{q}^2 + \overline{m}^2} \right)\right],
\end{aligned}
\ee
\be
\begin{aligned}
f^{SD}\left(\overline{q}, \overline{k}\right) = -\frac{1}{2\pi} \int_{-1/2}^{1/2} & d\beta  \frac{1}{ \sqrt{ \overline{k}^2 - 2\beta\, \overline{k} \cdot \overline{q} + \overline{q}^2 } } \\
& \cdot \left[ \frac{1}{\sqrt{\overline{k}^2-2\beta\, \overline{k} \cdot \overline{q} + \overline{q}^2}} - \frac{ 1 + \overline{m}^2 + \overline{k}^2 - 2\beta\, \overline{k} \cdot \overline{q} + \frac{1}{4} \overline{q}^2 }{ \overline{k}^2 - 2\beta\, \overline{k} \cdot \overline{q} + \overline{q}^2} \right] \\
& \cdot \ln \left[ \frac{ 1 + 2\sqrt{ \overline{k}^2 - 2\beta\, \overline{k} \cdot \overline{q} + \beta^2 \overline{q}^2 } + \overline{k}^2 - 2\beta\, \overline{k} \cdot \overline{q} + \frac{1}{4} \overline{q}^2 + \overline{m}^2 }{ 1 - 2 \sqrt{ \overline{k}^2 - 2\beta\, \overline{k} \cdot \overline{q} + \beta^2 \overline{q}^2 } + \overline{k}^2 - 2\beta\, \overline{k} \cdot \overline{q} + \frac{1}{4} \overline{q}^2 + \overline{m}^2 } \right].
\end{aligned}
\ee
To determine the Fermi momentum of ${}_{13}^{27}\textrm{Al}$, we linearly interpolate between the experimentally measured Fermi momenta of ${}_{12}^{24}\textrm{Mg}$ and ${}_{20}^{40}\textrm{Ca}$ \cite{Moniz1971}. This results in a Fermi momentum of $K_F = 238\pm5 \, \textrm{MeV}$.

Note that $f^{SI}$ and $f^{SD}$ have an angular dependence due to the presence of $\overline{k}\cdot\overline{q}$. However, $f^{SI}$ and $f^{SD}$ do not vary significantly over the range of possible angular values. As such, the functions can be averaged over all angular values so that they effectively only depend on the magnitude of $\overline{k}$ and $\overline{q}$. Furthermore, it is expected that $\left| \overline{q} \right| \approx \frac{m_\mu}{K_F}$ because the process of interest is coherent $\mu-e$ conversion. Thus, $f^{SI}$ and $f^{SD}$ are effectively only functions of $\left|\overline{k}\right|$.

\begin{figure}[h]
	\subfloat[Probability distribution of the magnitude of nucleon momenta in ${}^{27}$Al as a function of the dimensionless momentum.]{
		\includegraphics[width=.45\textwidth]{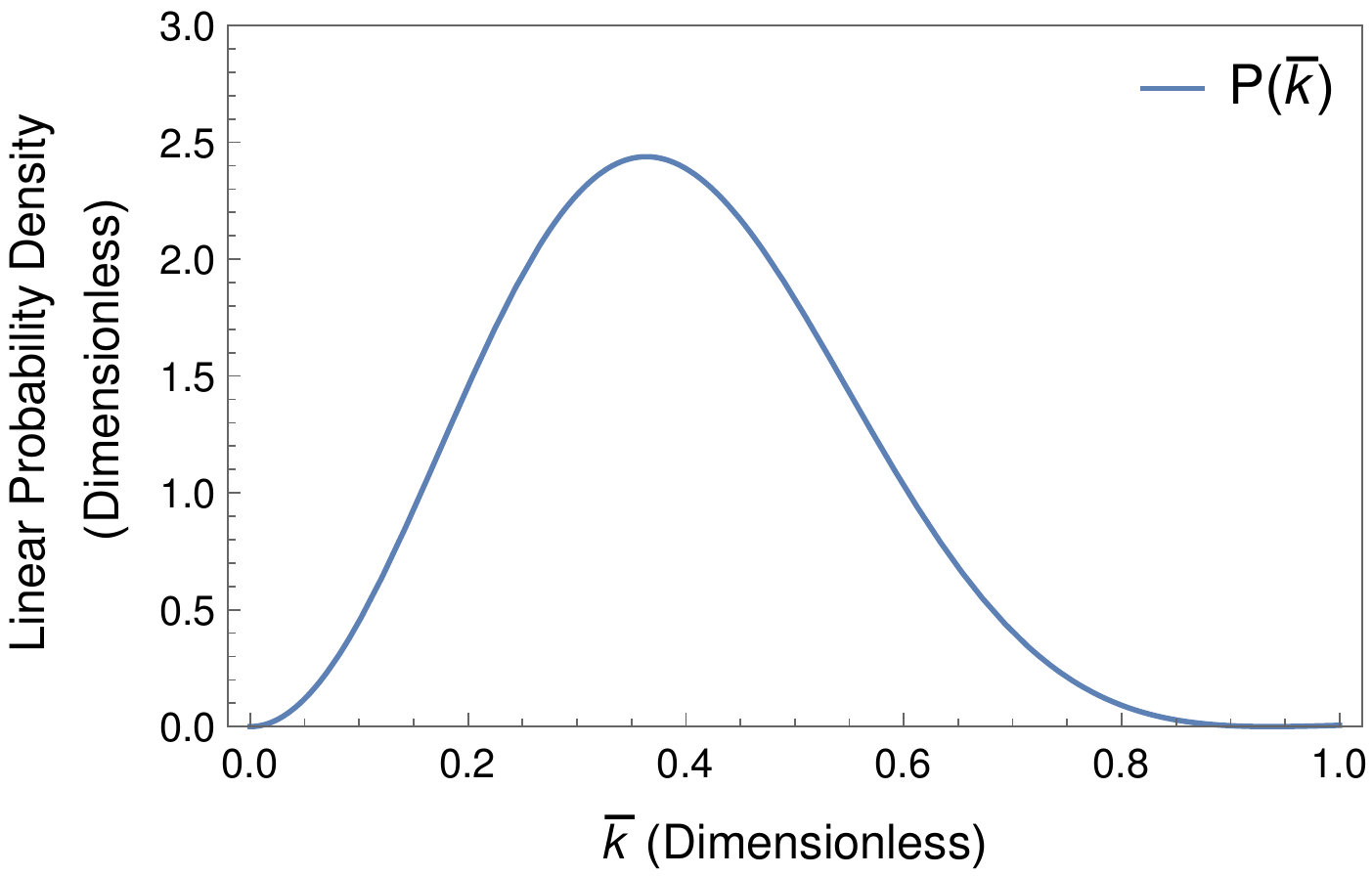}\label{fig:Core-Momenta}
		}\\
	\subfloat[The angle averaged value of $f^{SI}$ and its constant approximation $f^{SI}_{\textrm{eff}}$ as a function of the dimensionless average momentum.]{    			
		\includegraphics[width=.45\textwidth]{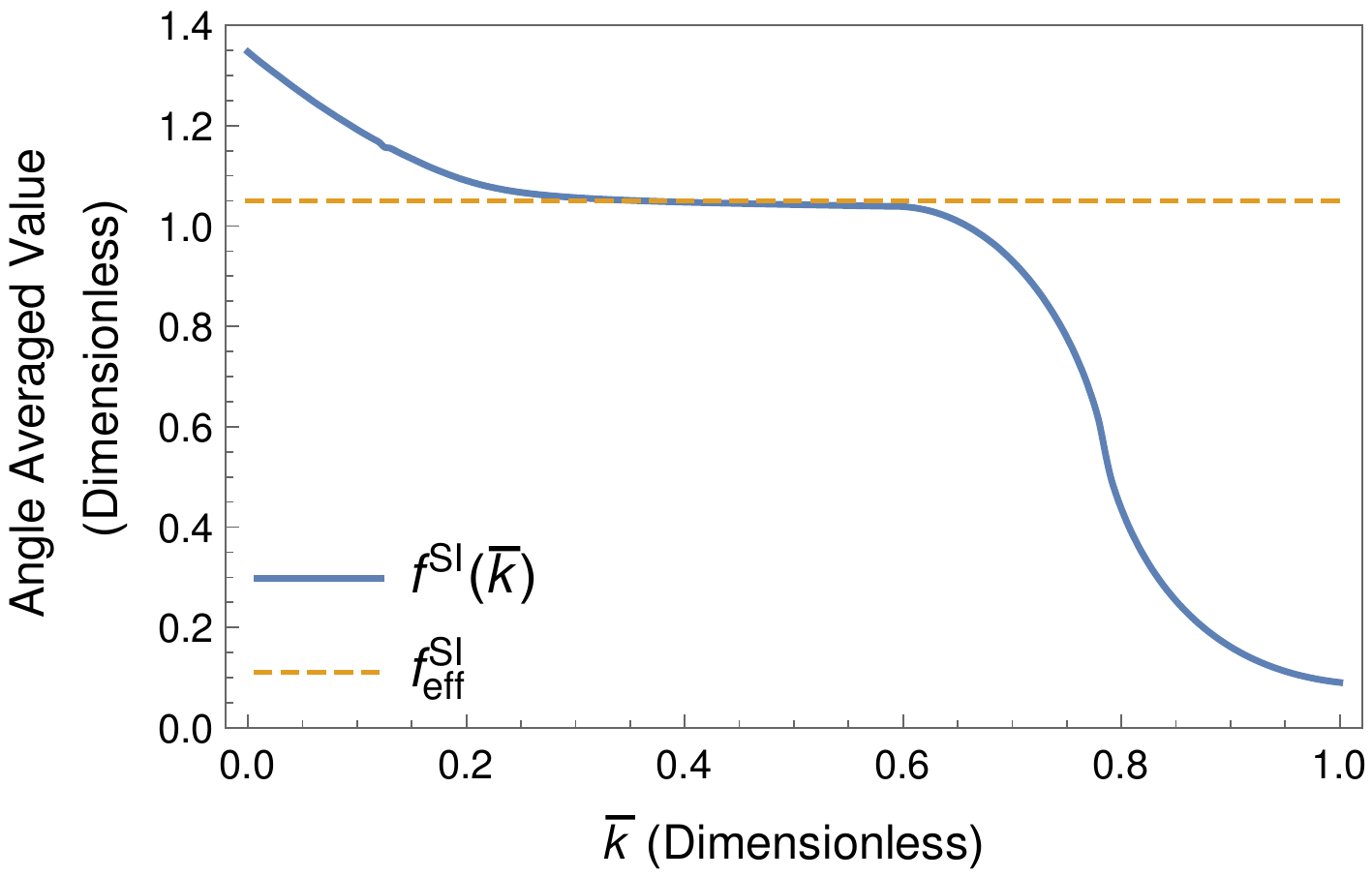}\label{fig:fSI-dependence}
		} \quad
	\subfloat[The angle averaged value of $f^{SD}$ and its constant approximation 						$f^{SD}_{\textrm{eff}}$ as a function of the dimensionless 								average momentum.]{
		\includegraphics[width=.45\textwidth]{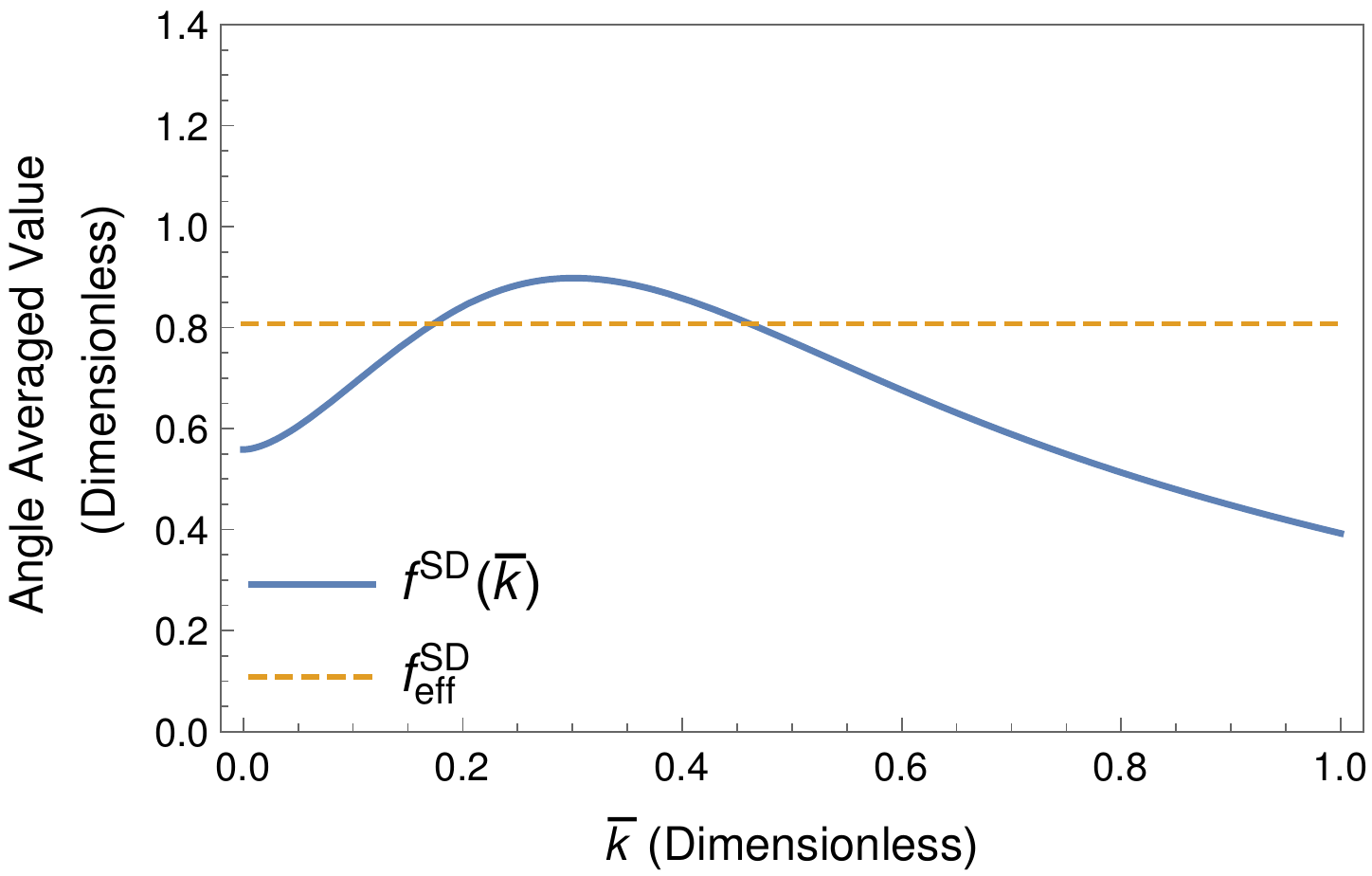}\label{fig:fSD-dependence}
		}
	\caption{Momentum dependence of the functions $f^{SI}$ and $f^{SD}$}
\end{figure}

Importantly, the momentum dependence of $f^{SI}$ and $f^{SD}$ is only of interest over the range of momenta common for nucleons in ${}^{27}$Al. Using the model-independent Fourier-Bessel expansion of the proton density distribution \cite{deVries1987}, the corresponding momentum distribution is shown in Figure \ref{fig:Core-Momenta}. It is important to note that Figure \ref{fig:Core-Momenta} is a plot of the linear probability density which integrates to unity with respect to $d\left|\overline{k}\right|$.

Figures \ref{fig:fSI-dependence} and \ref{fig:fSD-dependence} show $f^{SI}$ and $f^{SD}$ respectively over the same range of momenta with $K_F = 238 \, \textrm{MeV}$. With the goal of arriving at a local effective operator in position space, it is necessary to approximate $f^{SI}$ and $f^{SD}$ by constants independent of the nucleon momentum. These constants, $f^{SI}_{\textrm{eff}}$ and $f^{SD}_{\textrm{eff}}$, are chosen to minimize the weighted RMS error with respect to $f^{SI}$ and $f^{SD}$. The RMS weights are given by the nucleon momentum distribution. Taking into account both the RMS error and uncertainty in the Fermi momentum, we find $f^{SI}_{\textrm{eff}}=1.05\pm0.07$ and $f^{SD}_{\textrm{eff}}=0.81\pm0.12$.

\newpage
\section{Nuclear Density Parameters}
\label{app:Nuclear}

The proton density distribution of a nucleus can be parameterized in a model-independent manner using the Fourier-Bessel expansion
\be
\rho_{p}\left(r\right) = 
\begin{cases}
\sum_{n} a_{n} j_{0}\left(\frac{n \pi r}{R}\right) & r \leq R\\
0 & r > R
\end{cases}.
\ee
This distribution is normalized such that $\int 4\pi r^{2} \rho\left(r\right)dr = Z$ and depends on the cutoff radius, $R$, and the magnitude of the various components, $a_n$. The values for ${}^{27}\textrm{Al}$ are given in Table \ref{tab:ProtonConst} as determined by electron scattering experiments \cite{deVries1987}.

The neutron density distribution of a nucleus is usually given in terms of the two-parameter Fermi distribution,
\be
\rho_{n}\left(r\right) = \frac{\rho_0}{1+e^{\frac{r-c}{z}}}.
\ee
The normalization factor $\rho_0$ is chosen such that $\int 4\pi r^{2} \rho\left(r\right)dr = A-Z$ while the thickness parameter, $z$, and radial parameter, $c$ describe the shape of the distribution. The values of these parameters for ${}^{27}\textrm{Al}$ are given in Table \ref{tab:NeutronConst} where the experimental and systematic uncertainties have been combined \cite{Garcia1992}.

\begin{table}[h]
    \begin{minipage}[t]{.45\linewidth}
         	\caption{Parameters for proton density distribution\label{tab:ProtonConst}}
	\begin{tabular}{l|D{,}{\cdot}{-1}}
	Quantity & \multicolumn{1}{l}{Accepted Value} \\ \hline
	$\left\langle r^{2} \right\rangle^{1/2}_{p} \, \left[ \textrm{fm.}\right]$ & \multicolumn{1}{l}{$3.035\pm 0.002$} \\ \hline
	$R \, \left[ \textrm{fm.}\right]$ & 7.0 \\
	$a_{1} \, \left[\textrm{fm.}^{-3}\right]$ & 4.3418,10^{-1} \\
	$a_{2} \, \left[\textrm{fm.}^{-3}\right]$ & 6.0298,10^{-1} \\
	$a_{3} \, \left[\textrm{fm.}^{-3}\right]$ & 2.8950,10^{-2} \\
	$a_{4} \, \left[\textrm{fm.}^{-3}\right]$ & -2.3522,10^{-1} \\
	$a_{5} \, \left[\textrm{fm.}^{-3}\right]$ & -7.9791,10^{-2} \\
	$a_{6} \, \left[\textrm{fm.}^{-3}\right]$ & 2.3010,10^{-2} \\
	$a_{7} \, \left[\textrm{fm.}^{-3}\right]$ & 1.0794,10^{-2} \\
	$a_{8} \, \left[\textrm{fm.}^{-3}\right]$ & 1.2574,10^{-3} \\
	$a_{9} \, \left[\textrm{fm.}^{-3}\right]$ & -1.3021,10^{-3} \\
	$a_{10} \, \left[\textrm{fm.}^{-3}\right]$ & 5.6563,10^{-4} \\
	$a_{11} \, \left[\textrm{fm.}^{-3}\right]$ & -1.8011,10^{-4} \\
	$a_{12} \, \left[\textrm{fm.}^{-3}\right]$ & 4.2869,10^{-5} \\
	\end{tabular}
    \end{minipage} \quad \quad
    \begin{minipage}[t]{.45\linewidth}
    	\caption{Parameters for neutron density distribution \label{tab:NeutronConst}}
	\begin{tabular}{l|l}
	Quantity & Accepted Value \\ \hline
	$\left\langle r^{2} \right\rangle^{1/2}_{n} \, \left[ \textrm{fm.}\right]$ & $3.17\pm 0.11$ \\ \hline
	$c \, \left[ \textrm{fm.}\right]$ & $3.18\pm0.19$ \\
	$z \, \left[\textrm{fm.}\right]$ & $0.535$ \\
	\end{tabular}
    \end{minipage} 
\end{table}

\newpage
\section{Model Independent Overlap Integrals}
\label{app:Overlap}

In Section \ref{sec:BranchingRatio}, the coherent conversion rate \eqref{eq:SimpleDecayRate} was expressed as a sum of transition probabilities over eight possible spin configurations. However, there is a two-fold symmetry in the choice of overall sign for the spins. Thus, there are only four independent configurations. For compactness of notation, an index $w\in\left\{1,2,3,4\right\}$ is used to denote each unique configuration. The relationship between all possible spin configurations and $w$ is given in Table \ref{tab:SpinConfig}.

For a fixed spin configuration, one can perform the position space integral over the leptonic part of \eqref{eq:SimpleDecayAmplitude} to arrive at a function of the momentum transfer. For the case of scalar-mediated conversion, these are given by the dimensionless functions
\begin{align}
Z_{S}^{\left(1\right)}\left(\left|q_T\right|\right) &= \sqrt{m_{\mu}} \int dx \frac{1}{2\pi^2} \left| x \right|^2 j_0 \left( \left| x \right| \left| q_T \right| \right) \left( g_{-1}^{(e)}\left(x\right) g_{-1}^{(\mu)}\left(x\right) + f_{-1}^{(e)}\left(x\right) f_{-1}^{(\mu)}\left(x\right) \right), \\
Z_{S}^{\left(2\right)}\left(\left|q_T\right|\right) &= \sqrt{m_{\mu}} \int dx \frac{1}{8\pi} \left| x \right|^2 j_1 \left( \left| x \right| \left| q_T \right| \right) \left( g_{-1}^{(e)}\left(x\right) f_{-1}^{(\mu)}\left(x\right) - f_{-1}^{(e)}\left(x\right) g_{-1}^{(\mu)}\left(x\right) \right), \\
Z_{S}^{\left(3\right)}\left(\left|q_T\right|\right) &= \sqrt{m_{\mu}} \int dx \frac{1}{8\pi} \left| x \right|^2 j_1 \left( \left| x \right| \left| q_T \right| \right) \left( g_{+1}^{(e)}\left(x\right) g_{-1}^{(\mu)}\left(x\right) + f_{+1}^{(e)}\left(x\right) f_{-1}^{(\mu)}\left(x\right) \right), \\
Z_{S}^{\left(4\right)}\left(\left|q_T\right|\right) &= \sqrt{m_{\mu}} \int dx \frac{1}{2\pi^2} \left| x \right|^2 j_0 \left( \left| x \right| \left| q_T \right| \right) \left( f_{+1}^{(e)}\left(x\right) g_{-1}^{(\mu)}\left(x\right) - g_{+1}^{(e)}\left(x\right) f_{-1}^{(\mu)}\left(x\right) \right).
\end{align}

One can then perform the remaining momentum integrals of \eqref{eq:SimpleDecayAmplitude} in a model-independent manner. The only CLFV operator that depends on momentum transfer is the $\textrm{arccot}$ term in the NLO loop contribution of \eqref{eq:scalar-full-eff-Lagrangian}. This term will be associated with the overlap integral $\tilde{I}_{S,\alpha}^{\left(w\right)}$. All other CLFV operators are independent of momentum transfer and will be accompanied by the overlap integral $I_{S,\alpha}^{\left(w\right)}$. Defining the Fourier transformed nucleon density as $\widetilde{\rho}_{\alpha} \left( k \right) = \widetilde{\psi}^{*}_{\alpha} \left( k \right) \widetilde{\psi}_{\alpha} \left( k \right)$, the definitions for these two overlap integrals are given by
\begin{align}
& I_{S,\alpha}^{\left(w\right)} = \frac{1}{m_{\mu}^{5/2}}\int d q_{T} \int d q_{A} \left| q_{T} \right|^2 \left| q_{A} \right|^2 Z_{S}^{\left(w\right)} \left(\left|q_T\right|\right) \widetilde{\rho}_{\alpha} \left( \frac{1}{2} \sqrt{\left| q_{T} \right|^2 + \left| q_{A} \right|^2} \right), \\
& \begin{multlined}[t][.87\textwidth]
\tilde{I}_{S,\alpha}^{\left(w\right)} = \frac{1}{m_{\mu}^{5/2}} \int d q_{T} \int d q_{A} \left| q_{T} \right|^2 \left| q_{A} \right|^2 Z_{S}^{\left(w\right)} \left(\left|q_T\right|\right) \widetilde{\rho}_{\alpha} \left( \frac{1}{2} \sqrt{\left| q_{T} \right|^2 + \left| q_{A} \right|^2} \right) \\
\cdot \frac{2+X_{\pi}}{\sqrt{X_{\pi}}} \textrm{arccot} \left( \frac{2}{\sqrt{X_{\pi}}} \right).
\end{multlined}
\end{align}

In the case of vector-mediated conversion, one instead finds that the leptonic part of \eqref{eq:SimpleDecayAmplitude} can be reduced to the functions
\begin{align}
Z_{V}^{\left(1\right)}\left(\left|q_T\right|\right) &= \sqrt{m_{\mu}} \int dx \frac{1}{2\pi^2} \left| x \right|^2 j_0 \left( \left| x \right| \left| q_T \right| \right) \left( g_{-1}^{(e)}\left(x\right) g_{-1}^{(\mu)}\left(x\right) - f_{-1}^{(e)}\left(x\right) f_{-1}^{(\mu)}\left(x\right) \right), \\
Z_{V}^{\left(2\right)}\left(\left|q_T\right|\right) &= \sqrt{m_{\mu}} \int dx \frac{1}{8\pi} \left| x \right|^2 j_1 \left( \left| x \right| \left| q_T \right| \right) \left( g_{-1}^{(e)}\left(x\right) f_{-1}^{(\mu)}\left(x\right) + f_{-1}^{(e)}\left(x\right) g_{-1}^{(\mu)}\left(x\right) \right), \\
Z_{V}^{\left(3\right)}\left(\left|q_T\right|\right) &= \sqrt{m_{\mu}} \int dx \frac{1}{8\pi} \left| x \right|^2 j_1 \left( \left| x \right| \left| q_T \right| \right) \left( g_{+1}^{(e)}\left(x\right) g_{-1}^{(\mu)}\left(x\right) - f_{+1}^{(e)}\left(x\right) f_{-1}^{(\mu)}\left(x\right) \right), \\
Z_{V}^{\left(4\right)}\left(\left|q_T\right|\right) &= - \sqrt{m_{\mu}} \int dx \frac{1}{2\pi^2} \left| x \right|^2 j_0 \left( \left| x \right| \left| q_T \right| \right) \left( f_{+1}^{(e)}\left(x\right) g_{-1}^{(\mu)}\left(x\right) + g_{+1}^{(e)}\left(x\right) f_{-1}^{(\mu)}\left(x\right) \right).
\end{align}

Unlike the case of scalar-mediated CLFV, no term in the vector-mediated CLFV Lagrangian depends on the momentum transfer. The only type of overlap integral is
\be
I_{V,\alpha}^{\left(w\right)} = \frac{1}{m_{\mu}^{5/2}} \int d q_{T} \int d q_{A} \left| q_{T} \right|^2 \left| q_{A} \right|^2 Z_{V}^{\left(w\right)} \left(\left|q_T\right|\right) \widetilde{\rho}_{\alpha} \left( \frac{1}{2} \sqrt{\left| q_{T} \right|^2 + \left| q_{A} \right|^2} \right).
\ee

The numerical values of $I_{S,\alpha}^{\left(w\right)}$, $\tilde{I}_{S,\alpha}^{\left(w\right)}$, and $I_{V,\alpha}^{\left(w\right)}$ can readily be calculated using the proton and neutron distributions of Appendix \ref{app:Nuclear} along with the muon and electron wavefunctions calculated from them. The values of $I_{S,\alpha}^{\left(w\right)}$, $\tilde{I}_{S,\alpha}^{\left(w\right)}$, and $I_{V,\alpha}^{\left(w\right)}$ along with their uncertainties are cataloged in Table \ref{tab:Overlap}.

\begin{table}[h]
\centering
\begin{tabular}{r|r|r|r||r}
$\kappa_{i}$ & $\mu_{i}$ & $\kappa_{f}$ & $\mu_{f}$ & $w$ \\ \hline
-1 & -$\frac{1}{2}$ & -1 & -$\frac{1}{2}$ & 1\\
-1 & +$\frac{1}{2}$ & -1 & +$\frac{1}{2}$ & 1\\
-1 & -$\frac{1}{2}$ & -1 & +$\frac{1}{2}$ & 2\\
-1 & +$\frac{1}{2}$ & -1 & -$\frac{1}{2}$ & 2\\
-1 & -$\frac{1}{2}$ & +1 & +$\frac{1}{2}$ & 3\\
-1 & +$\frac{1}{2}$ & +1 & -$\frac{1}{2}$ & 3\\
-1 & -$\frac{1}{2}$ & +1 & -$\frac{1}{2}$ & 4\\
-1 & +$\frac{1}{2}$ & +1 & +$\frac{1}{2}$ & 4
\end{tabular}
\caption{Table of spin configurations \label{tab:SpinConfig}}
\end{table}

\begin{table}[h]
\centering
\begin{tabular}{l|D{.}{.}{3,3}|l|D{,}{\pm}{6,3}}
Proton Overlap Integral & \multicolumn{1}{c|}{Value} & Neutron Overlap Integral & \multicolumn{1}{c}{Value} \\ \hline
$I^{1}_{S,p}$ & 7.58 & $I^{1}_{S,n}$ & 7.58, 0.24\\
$I^{2}_{S,p}$ & 5.50 & $I^{2}_{S,n}$ & 5.49, 0.17\\
$I^{3}_{S,p}$ & -5.53 & $I^{3}_{S,n}$ & -5.52, 0.17\\
$I^{4}_{S,p}$ & -7.56 & $I^{4}_{S,n}$ & -7.55, 0.24\\ \hline
$\tilde{I}^{1}_{S,p}$ & 9.57 & $\tilde{I}^{1}_{S,n}$ & 9.55, 0.31\\
$\tilde{I}^{2}_{S,p}$ & 6.93 & $\tilde{I}^{2}_{S,n}$ & 6.92, 0.23\\
$\tilde{I}^{3}_{S,p}$ & -6.96 & $\tilde{I}^{3}_{S,n}$ & -6.96, 0.23\\
$\tilde{I}^{4}_{S,p}$ & -9.52 & $\tilde{I}^{4}_{S,n}$ & -9.51, 0.30\\ \hline
$I^{1}_{V,p}$ & 7.35 & $I^{1}_{V,n}$ & 7.32, 0.24\\
$I^{2}_{V,p}$ & -5.79 & $I^{2}_{V,n}$ & -5.76, 0.19\\
$I^{3}_{V,p}$ & -5.81 & $I^{3}_{V,n}$ & -5.79, 0.19\\
$I^{4}_{V,p}$ & 7.31 & $I^{4}_{V,n}$ & 7.29, 0.24\\
\end{tabular}
\caption{Table of overlap integrals \label{tab:Overlap}}
\end{table}

\newpage
\section{Formula for the Branching Ratio}
\label{app:BR-Formula}

Given a CLFV Lagrangian of the form \eqref{eq:CLFV-Lagrangian-generic}, one can define the Wilson coefficients \eqref{eq:eff-lep-coef-scalar}-\eqref{eq:eff-lep-coef-mass}. As explained in Section \ref{sec:BranchingRatio}, the branching ratio for coherent $\mu-e$ conversion can be written as a sum over separate amplitudes for each spin configuration, \eqref{eq:SimpleDecayRate}. Accounting for symmetry in the spin configurations, the index $w\in\left\{1,2,3,4\right\}$ indicates the four independent spin configurations of the system as outlined in Table \ref{tab:SpinConfig} of Appendix \ref{app:Overlap}. Written as a sum over these four independent configuration, this yields the master equation for the branching ratio, \eqref{eq:brmaster}.

Each conversion amplitude for a specific spin configuration can then be expressed in terms of Wilson coefficients and a set of model-independent parameters. This is done for scalar-mediated conversion in \eqref{eq:ConvAmpl-Scalar} and for vector-mediated conversion in \eqref{eq:ConvAmpl-Vector}. The only model dependent-parameters that appear in these expressions are the Wilson coefficients; all model-independent parameters have been absorbed into the definitions of the $\alpha$'s. Using the definition of $\Delta_S^{(w)}$ from \eqref{eq:definition-delta}, these are defined as
\begin{align}
&\begin{multlined}[t][.87\textwidth]
\alpha_{S,ud}^{(w)} = \sqrt{ \frac{m_{\mu}}{\omega_{\textrm{capt}}} } \left( \frac{m_{\mu}}{4 \pi v} \right)^2 \left[ \frac{\sigma_{\pi N}}{2 \hat{m}} \left( I^{(w)}_{S,p} + I^{(w)}_{S,n} \right) - \frac{3B_{0}K_{F}\mathring{g}_{A}^{2}}{64\pi \mathring{f}_{\pi}^{2}} f^{SI}_{\textrm{eff}}\left( I^{(w)}_{S,p} + I^{(w)}_{S,n} \right) \right. \\
\left. - \frac{3B_{0}m_{\pi}\mathring{g}_{A}^{2}}{64\pi \mathring{f}_{\pi}^{2}} \Delta_S^{(w)} \right] ,
\end{multlined} \label{eq:a-ScalarUpDown} \\
&\alpha_{S,s}^{(w)} = \sqrt{ \frac{m_{\mu}}{\omega_{\textrm{capt}}} } \left( \frac{m_{\mu}}{4 \pi v} \right)^2 \frac{\sigma_{s N}}{m_{s}} \left( I^{(w)}_{S,p} + I^{(w)}_{S,n} \right) , \label{eq:a-ScalarStrange}\\
&\alpha_{S,\Theta}^{(w)} = \sqrt{ \frac{m_{\mu}}{\omega_{\textrm{capt}}} } \left( \frac{m_{\mu}}{4 \pi v} \right)^2 \left( I^{(w)}_{S,p} + I^{(w)}_{S,n} \right) , \label{eq:a-ScalarTheta}\\
&\alpha_{V,u}^{(w)} = \sqrt{ \frac{m_{\mu}}{\omega_{\textrm{capt}}} } \left( \frac{m_{\mu}}{4 \pi v} \right)^2 \left( 2 I^{(w)}_{V,p} + I^{(w)}_{V,n} \right) , \label{eq:a-VectorUp} \\
&\alpha_{V,d}^{(w)} = \sqrt{ \frac{m_{\mu}}{\omega_{\textrm{capt}}} } \left( \frac{m_{\mu}}{4 \pi v} \right)^2 \left( I^{(w)}_{V,p} + 2 I^{(w)}_{V,n} \right) . \label{eq:a-VectorDown}
\end{align}
The quantities $I^{(w)}_{S,x}$, $\tilde{I}^{(w)}_{S,x}$, and $I^{(w)}_{V,x}$ are the overlap integrals defined in Appendix \ref{app:Overlap} and given numerically in Table \ref{tab:Overlap}. The quantity $f^{SI}_{\textrm{eff}}=1.05_{-0.53}^{+0.07}$ characterizes the effective one-nucleon operator which is discussed in Section \ref{sec:Effective-One-Nucleon} and Appendix \ref{app:OneBody}. The remaining physical constants are given in Table \ref{tab:Constants} of Appendix \ref{app:Constants}.

As the $\alpha$ parameters are model-independent, they can be calculated in advance and their numerical values and uncertainties are given in Tables \ref{tab:BR-Scalar} and \ref{tab:BR-Vector} of Section \ref{sec:Introduction}. In the case of scalar-mediated conversion, the LO contributions and those of the loop diagram and two-nucleon diagram that enter at NLO can be analyzed separately.

\end{appendix}

%%%%%%%%%%%%%%%%%%%%%%%%%%%%%%%%%%%%%%%%%%%%%%%%%

\newpage

\bibliographystyle{utphys}
\bibliography{LFV}

\end{document}